\documentclass[12pt,preprint]{aastex}

\begin{document}

\title{A Spectroscopic Study of Young Stellar Objects in the Serpens Cloud Core and NGC 1333.}

\shorttitle{A Spectroscopic Survey of Serpens \& NGC 1333}

\author{E. Winston\altaffilmark{1,2}, S. T. Megeath\altaffilmark{1,3}, S. J. Wolk\altaffilmark{1}, J. Hernandez\altaffilmark{4,5} , R. Gutermuth\altaffilmark{1}, J. Muzerolle\altaffilmark{6}, J. L. Hora\altaffilmark{1} , K. Covey\altaffilmark{1}, L.E. Allen\altaffilmark{1}, B. Spitzbart\altaffilmark{1}, D. Peterson\altaffilmark{1}, P. Myers\altaffilmark{1}, G. G. Fazio\altaffilmark{1}}

\altaffiltext{1}{Harvard Smithsonian Center for Astrophysics, 60 Garden St., Cambridge MA 02138, USA.} 
\email{ewinston@cfa.harvard.edu}
\altaffiltext{2}{School of Physics, Science Centre - North, University College Dublin, Belfield, Dublin 4, Ireland.}
\altaffiltext{3}{Current address: Ritter Observatory, Dept. of Physics and Astronomy, University of Toledo, 2801 W. Bancroft Ave., Toledo, OH 43606, USA. }
\altaffiltext{4}{Centro de Investigaciones de Astronomia, Apdo. Postal 264, Merida 5101-A, Venezuela.  }
\altaffiltext{5}{University of Michigan, Ann Arbor, MI 48109.   }
\altaffiltext{6}{Steward Observatory, University of Arizona, 933 N. Cherry Ave., Tucson, AZ 85721. }

\begin{abstract}

We present spectral observations of 130 young stellar objects (YSOs) in the 
Serpens Cloud Core and NGC 1333 embedded clusters.   The observations 
consist of near-IR spectra in the $H$ and $K$-bands, from SpeX on the IRTF 
and far-red spectra (6000 - 9000~\AA)  from Hectospec on the MMT.  
These YSOs were identified in previous {\it Spitzer} and {\it Chandra} observations, 
and the evolutionary classes of the YSOs were determined from the {\it Spitzer} 
mid-IR photometry.  
With these spectra, we search for corroborating evidence for the pre-main 
sequence nature of the objects, study the properties of the detected emission 
lines as a function of evolutionary class, and obtain spectral types for the 
observed YSOs.  
The temperature implied by the spectral types are combined with luminosities 
determined from the near-IR photometry to construct HR diagrams for the clusters.  
By comparing the positions of the YSOs in the HR diagrams with the pre-main 
sequence tracks of \citet{bar}, we determine ages of the embedded sources and 
study the relative ages of the YSOs with and without optically thick circumstellar disks.   
The apparent isochronal ages of the YSOs in both clusters range from  
less than 1 Myr to 10 Myr, with most objects below 3 Myr.
The observed distribution of ages for the Class II and Class III  
objects are statistically indistinguishable.
We examine the spatial distribution and extinction of the YSOs as a  
function of their  isochronal ages.  We find the sources $<$ 3 Myr
to be concentrated in the molecular cloud gas while the older sources  
are spatially dispersed and are not deeply embedded.
Nonetheless, the sources with isochronal ages $>$ 3 Myr show all the  
characteristics of young stellar objects in their spectra, their IR  
spectral energy distributions, and their X-ray emission; we find no evidence  
that they are contaminating background giants or foreground dwarfs.  
However, we find no corresponding decrease in the fraction of sources with 
infrared excess with isochronal age; this suggests that the older isochronal 
ages may not measure the true age of the $>$ 3 Myr YSOs.
Thus, the nature of the apparently older sources and their  
implications for cluster formation remain unresolved.

\end{abstract}


\today

\section{\bf Introduction}

In this paper, we describe a detailed spectroscopic study of the population  
of young stellar objects (YSOs) in the Serpens and NGC 1333 clusters identified 
previously using observations  with {\it Spitzer} and {\it Chandra} by 
\citet{winston}, \citet{gutn}, \citet{win08b}.  
Spectroscopy can establish the pre-main sequence nature of candidate members of 
these regions, detect emission lines thought to result from accretion of gas onto 
YSOs, and determine spectral classifications for each of the members in order to 
construct HR diagrams.  The spectral class can be used in combination with 
near-IR photometry and pre-main sequence models to determine ages and masses for 
the YSOs. These measured ages and masses can be used to determine 
the initial mass functions of embedded clusters, study the star formation 
history of the clusters, and examine the evolution of circumstellar disks as a 
function of age.

The Serpens region is an example of a very young, deeply embedded cluster, 
containing a high percentage of protostars \citep{har,hur,eir1,eir2,kaa,tes2,winston}. 
The embedded cluster is heavily extinguished, with a peak extinction exceeding 40 
magnitudes in the visual.  
Infrared and (sub)millimeter observations of the regions have  
identified at least 38 protostars, including Class 0, I and
flat spectrum sources \citep{tes1,tes2,dav1,hog,winston}. 
On the basis of the K-band  magnitude distribution, \citet{kaa} proposed that the cluster  
contained two populations: one with an age of  $\sim1\times10^{5}$~yrs and another 
with an age  of $\sim2\times10^{6}$~yrs.
At a distance of $\sim$260~pc, the Serpens cloud core is one of the nearest regions 
of clustered star-formation to the Sun (see \citet{strai} for a discussion of the 
distance to Serpens). This makes it an excellent candidate for study  as it is 
close enough to both resolve the individual members and to detect the lowest mass 
members to below the hydrogen-burning limit.

The NGC 1333 region is another example of an embedded, relatively nearby young cluster.
The distance to the region has recently been measured as 240~pc \citep{hirota}. This 
distance is based on a VERA measurement of a maser parallax toward SVS 13, and is the 
distance used in this study.  Previous estimates range from 250~pc \citep{enoch} to 
340~pc \citep{cernis}, based on the {\it Hipparcos} parallax distance to the Perseus OB2 association.
The cluster contains a high concentration of molecular outflows and a large number of deeply 
embedded protostars, along with numerous cores detected in the sub-mm \citep{knee,jenn}.  
The pre-main sequence members have previously been identified through optical/near-IR surveys, 
number counts analysis \citep{strom,asp1,wilk,lada} and recent {\it Spitzer} work \citep{gutn}.

This paper presents spectroscopy of Spitzer and Chandra identified YSOs from two 
instruments: the far-red spectrograph Hectospec on the MMT, and the near-IR 
spectrograph SpeX on the IRTF.    
The sample consists primarily of pre-main sequence stars both with optically thick 
disks of gas and dust (Class II) and without detectable disks (Class III), as well as 
a few protostars (Class 0/I and Flat Spectrum sources).    
Given the range of extinctions toward the member YSOs of our target regions, 
we obtained spectra with two separate instruments covering different wavelength 
regimes.  
The multi-object far-red spectra sample obtained sample the more dispersed, less 
embedded YSOs as well as randomly selected stars in the 
field with similar colors and magnitudes. Some of these randomly selected stars 
may be Class III objects without IR-excesses or X-ray emission.   
The sources too embedded to detect in the far-red were subsequently observed using 
single-object near-IR spectroscopy.  
In this paper,  the spectral classifications of the identified YSOs in the Serpens 
and NGC 1333 clusters are utilized in creating Hertzsprung-Russell diagrams for 
the two clusters.  The isochronal ages are used to study the star formation history 
of these clusters and compare the ages of objects with and without IR-excesses due 
to optically thick disks.  This is used to study the apparent age spreads observed 
toward many embedded clusters.  
The spatial distribution and extinction of the YSOs are examined as a function of age.
In addition, a brief investigation of $H{\alpha}$ and other emission lines as a 
function of evolutionary class and spectral classification is presented.   
Since a complete spectroscopic sample of either cluster has not been achieved, the 
IMF is not discussed in this contribution.  
The observations and data reduction are described in \S~\ref{obsdata},  \S~\ref{results} 
details the results of the spectral classification and emission line studies, while the 
discussion of the isochronal ages is presented in \S~\ref{disc}.

\section{\bf Observations and Data Reduction}\label{obsdata}

\subsection{Far Red Spectroscopy: Hectospec}

Spectroscopy in the far-red was carried out using the Hectospec  multi-object moderate 
dispersion spectrograph  \citep{fab} on the converted 6.5~m Multi-Mirror Telescope (MMT) 
on Mt. Hopkins, Arizona.   
The spectral range covered was 3000-9000~\AA, with a resolution of $\sim$6~\AA.
Time was awarded on the MMT for three runs, consisting of one half night each,
in October 2005, March 2006, and November 2006. In total 395 sources were observed in
Serpens, and 382 sources in NGC 1333, of which the majority (86\%, 84\% respectively) were 
stars that had not previously been identified as YSOs in either the IR or X-rays.  
The total integration time for the individual spectra ranged from 300 to 600~seconds.  
The spectra were reduced in-house using the Telescope Data Center (TDC) Hectospec  
pipeline at the Harvard-Smithsonian CfA \citep{mink}. 
The IRAF-based pipeline extracted and calibrated the spectra.
Selected sky absorption features and a {\it red leak} near 9000~\AA~ are removed from the final spectra.

\subsection{Near IR Spectroscopy: SpeX}

Spectroscopy in the near-IR was carried out using the SpeX cryogenic spectrograph and 
imager on the 3.0~m NASA Infra-Red Telescope Facility (IRTF) on Mauna Kea, Hawai'i. 
The SpeX instrument was built by the Institute for Astronomy, University of Hawai'i \citep{rayn}. 
It is a medium resolution spectrograph, covering a wavelength range from 0.8-5.5~${\mu}m$, and 
employs prism cross-dispersion to provide simultaneous wavelength coverage.
The instrument was operated in SXD (0.8-2.4~${\mu}m$ Crossed-Dispersed) mode, 
providing wavelength coverage of 0.8-2.4~${\mu}m$, with slit widths of 0.3$''$ or 0.5$''$ 
(chosen depending on seeing conditions), and a slit length of 15$''$,  
with an image scale of 0.15$''$ pixel$^{-1}$ along the slit. 
Time was awarded on the IRTF for four runs of 3-4 nights each, on the 30 June - 2 July 2006 and 
June 28-July 1 2007 for observations of Serpens,  and the 12-14 of November 2006 and 26-29 
December 2007 for NGC 1333.

The SpeX spectra were reduced using SpeXtool, an IDL software package, specifically 
designed for the SpeX instrument \citep{cush,vacca}, which performs extraction, reduction and 
telluric correction.   
In the SXD mode, offsetting of the telescope along the slit provides a measure of the sky, 
and so all frames are taken in A \& B pairs.  The spectra are flat-fielded, extracted, 
wavelength calibrated, and sky-subtracted using calibration data taken between each exposure.  
Between each target exposure, a standard star was observed to provide a method of removing
residual telluric lines.   A single $A0V$ standard was observed in each cluster; HD 171149 in 
Serpens, and HD 22859 in NGC 1333.   Finally, the spectral orders are combined into a single 
spectrum, which is then cleaned to remove noise at the edges of the orders and cosmic ray hits.

\section{\bf Results}\label{results}

\subsection{Source Selection}

\subsubsection{Hectospec}
To identify objects that would have enough flux to be observed in the optical 
far-red given the strong extinction towards the regions, a  stringent near-IR 
color and magnitude criteria were adopted.   
Sources were required to be detected by 2MASS in the $J$ and $H$ -bands 
with uncertainties less than 0.2~mag and to satisfy the color and magnitude 
criteria of J-H $<$ 1.6 and J $<$ 15.  These criteria were based on the 
signal-to-noise of spectra from initial observations of Serpens, for which less 
stringent criteria were applied.   
To best utilize the 300 fibers available, we tried to obtain  as many of the 
{\it Spitzer} and {\it Chandra} identified YSOs as possible in a single fiber 
configuration;  however, due to fiber packing and positioning constraints, most 
of the fibers were not allocated to the identified YSOs.  The remaining fibers 
were instead randomly assigned to stars in the field which satisfied the color 
and magnitude criteria to serendipitously search for  Class III members not 
detected in the {\it Chandra} data.  These candidates were identified through 
the presence of $Li~I$ in absorption in their spectra, c.f. Sec.~\ref{lithium}.

\subsubsection{SpeX}

The SpeX observations solely observed YSOs previously identified either by their infrared-excesses 
in the $Spitzer$ bands or by their X-ray emission in the $Chandra$ observations \citep{winston,gutn,win08b}. 
The sources were classified with the $Spitzer$ photometry using the approach of \citet{winston,gutn}.
Of the 30 Serpens sources observed, 12 were Class II (10 of which were also detected in X-rays), 
12 were Class III (all 12 with X-rays), and 6 were transition disks (none detected with {\it Chandra}).  
In NGC 1333, 22 YSOs were observed, 13 were Class II (10 detected in X-rays) and 9 were Class III (7 in X-rays). 
These sources were chosen because they were more deeply embedded cluster members, 
which though bright (all had J-band magnitudes $<$ 14), were not in the J - H color range 
required to be selected for observation with Hectospec.  Sources with 
a $J - H$ $>$ 1.4 were selected, with some sources  selected to overlap with the 
Hectospec sample to provide a comparison between the spectral typing results at the two wavelengths.

\subsubsection{Sample Selection and Completeness of the SpeX and Hectospec Observations}\label{combsh}

\begin{figure}
\epsscale{1}
\plotone{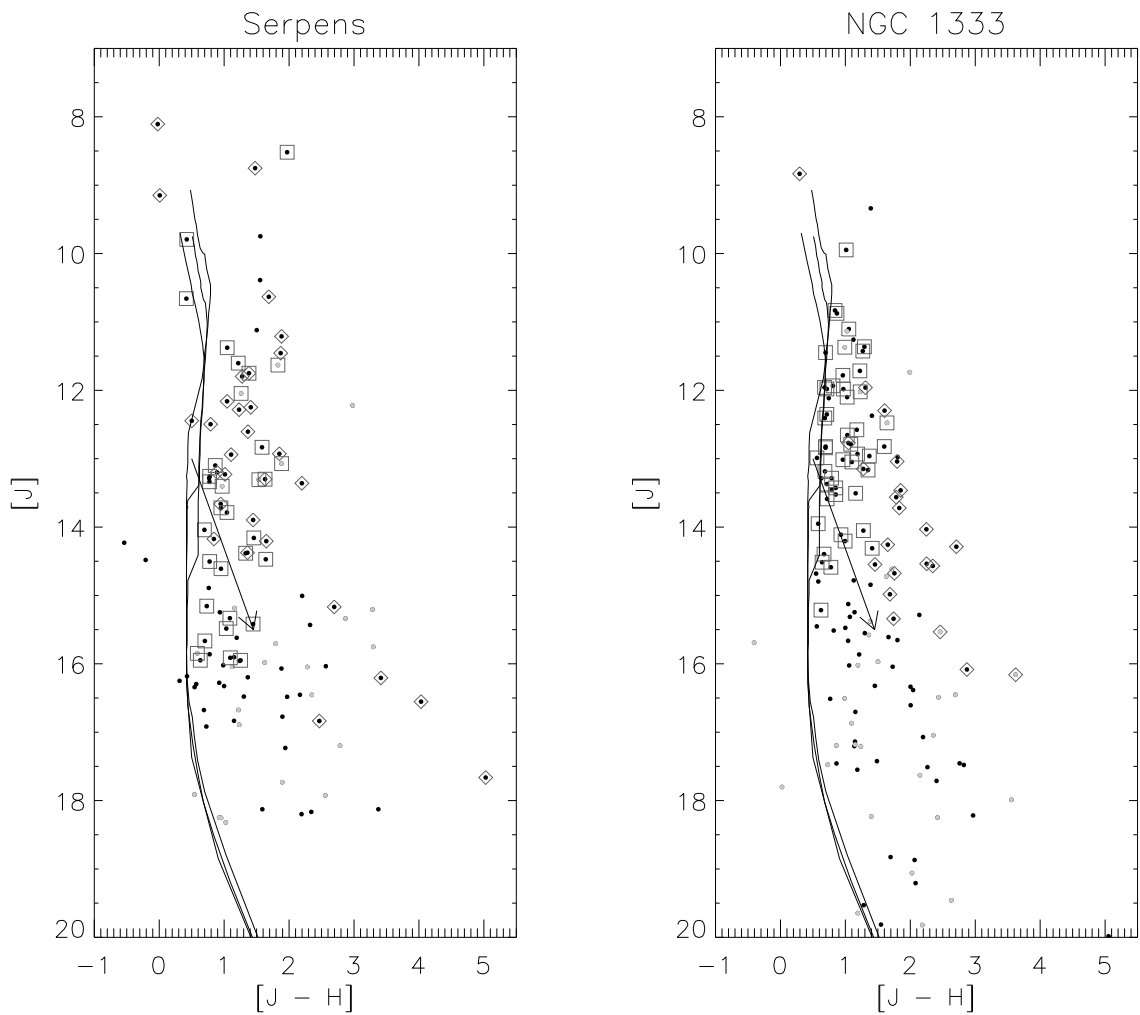}
\caption[Color-Magnitude of Spectral Sources]{ 
Color-Magnitude diagrams of the Serpens and NGC 1333 clusters, indicating the range in color and 
magnitude space covered by the spectral data sets in J-H \& J.  The small dots represent the 
known YSOs, with black dots showing the pre-main sequence stars, and the grey dots showing 
the protostellar objects. 
The black squares indicate those objects for which good spectral classifications were obtained with 
Hectospec, and the black diamonds indicate the SpeX sources.  
The \citet{bar} 1, 3, and 10~Myr isochrones are shown, as is a reddening vector of 1~$A_{K}$ (10~$A_{V}$).  
}
\label{cmdspec}
\end{figure}

Fig.~\ref{cmdspec} shows the color-magnitude plots for the two clusters, including all 
the previously identified YSOs and those observed in the two spectral observation sets, 
indicating the range in the near-IR  magnitude and color covered by the final spectral data 
set.  The spectrally classified objects were brighter than $J~=$ 16~mag. and had a $J - H$ 
color less than 4.5 mag.   
The Hydrogen-Burning Limit (HBL) at the distance of Serpens and NGC 1333 and an assumed age of 1~Myrs, 
is calculated from \citet{bar}, as $\sim$~13~mag at $J$-band, and is indicated by 
the placement of the reddening vector in Fig.~\ref{cmdspec}. 
Our samples are almost complete down to the HBL and to an $A_K$ of 1 ($A_V$ of 10):  only 4 known 
pre-main sequence members in this range lack spectra in Serpens, and 3 lack spectra in NGC 1333.  

\begin{figure}
\epsscale{0.8}
\plottwo{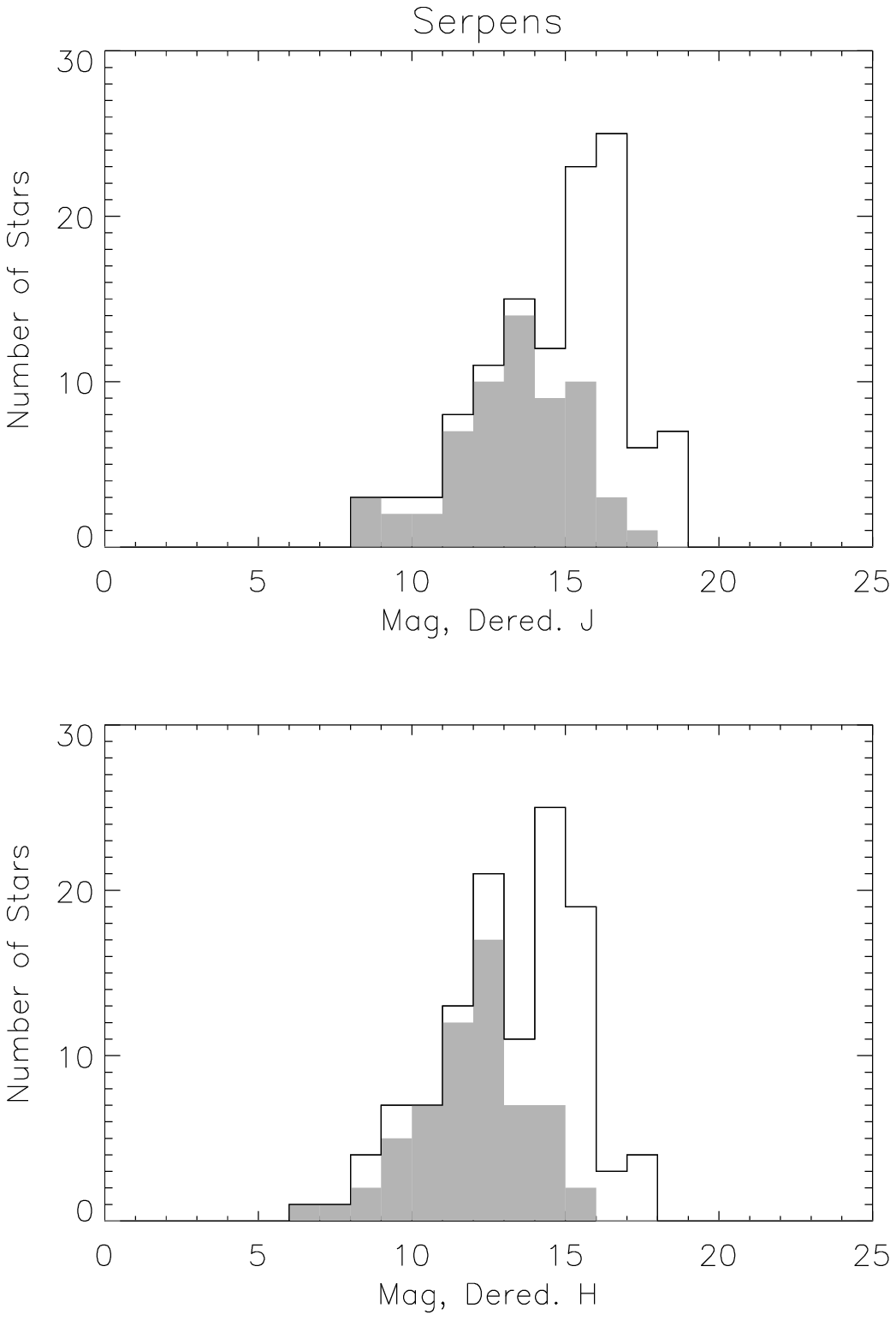}{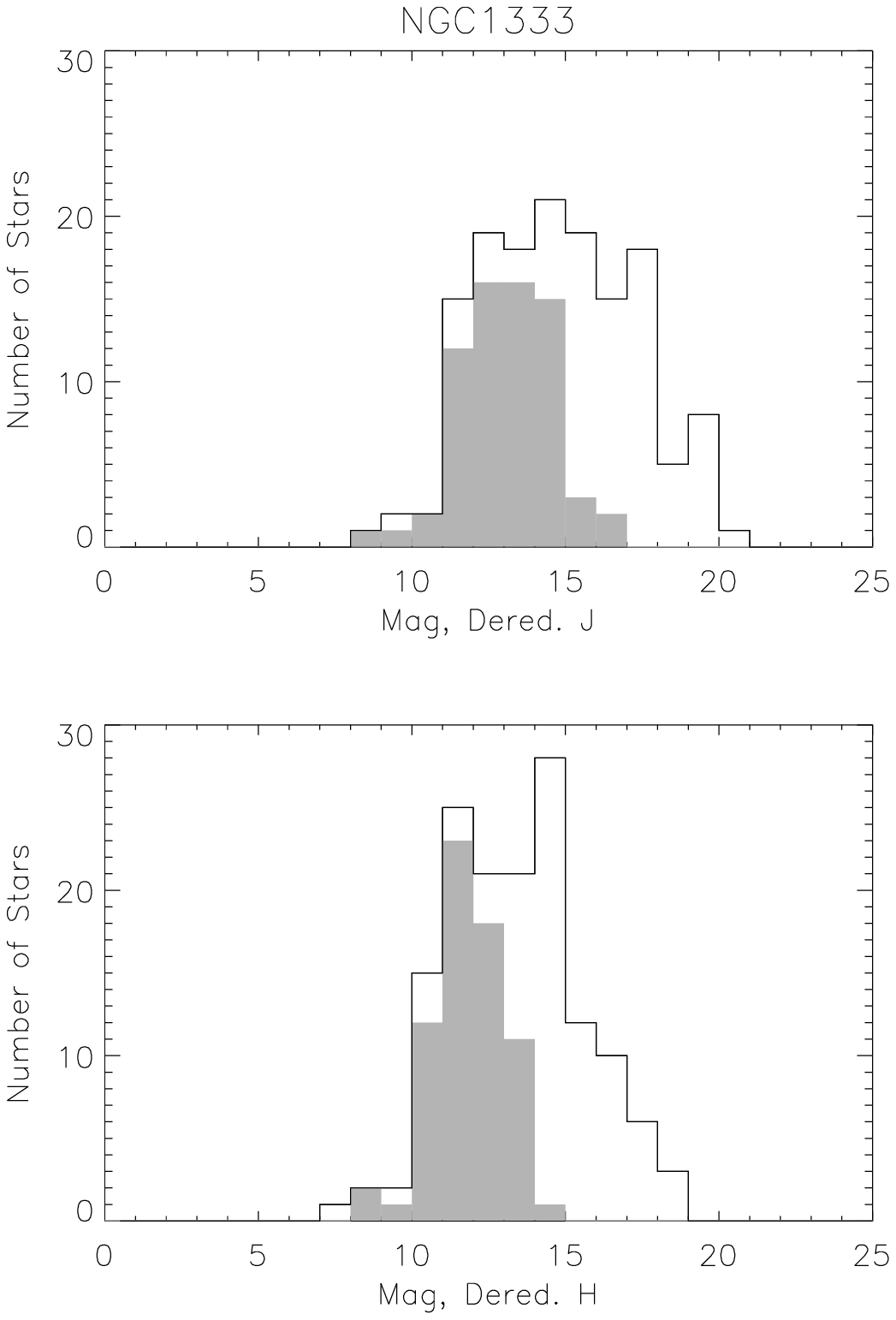}
\caption[Histogram of Dered. Magnitudes]{ 
Histograms of the dereddended $J$ and $H$-band magnitudes of the identified YSOs in 
the Serpens (left) and NGC 1333 (right) clusters.  
The solid black histograms are for all identified YSOs, the gray, filled-in histograms are for YSOs 
with spectral types.  The steep cutoff at J=15 in the filled-in histogram is due to our 
selection criteria for the spectroscopic sample. 
}
\label{histo_mags}
\end{figure}

In Serpens, combining the optical and near-IR observations, we have obtained spectral classifications 
for 61 of the 138 YSOs (45\%) previously identified in \citet{winston}. Of these, 35 have X-ray detections.
Likewise, in NGC 1333, 69 of the 169 YSOs (41\%) identified in \citet{gutn} were classified, of which, 
55 have X-ray detections, c.f. Table~\ref{ysonums}.   
In Figure~\ref{histo_mags}, the histograms of the dereddened $J$- and $H$-band magnitudes 
are plotted for all known cluster members and for the sub-sample of classified sources.  The dereddened 
magnitudes were determined following the method of \citet{gut1}, using the extinction law of \citet{fla}.

\subsection{Spectral Typing}

\subsubsection{Hectospec}

\begin{figure}
\epsscale{1.}
\plottwo{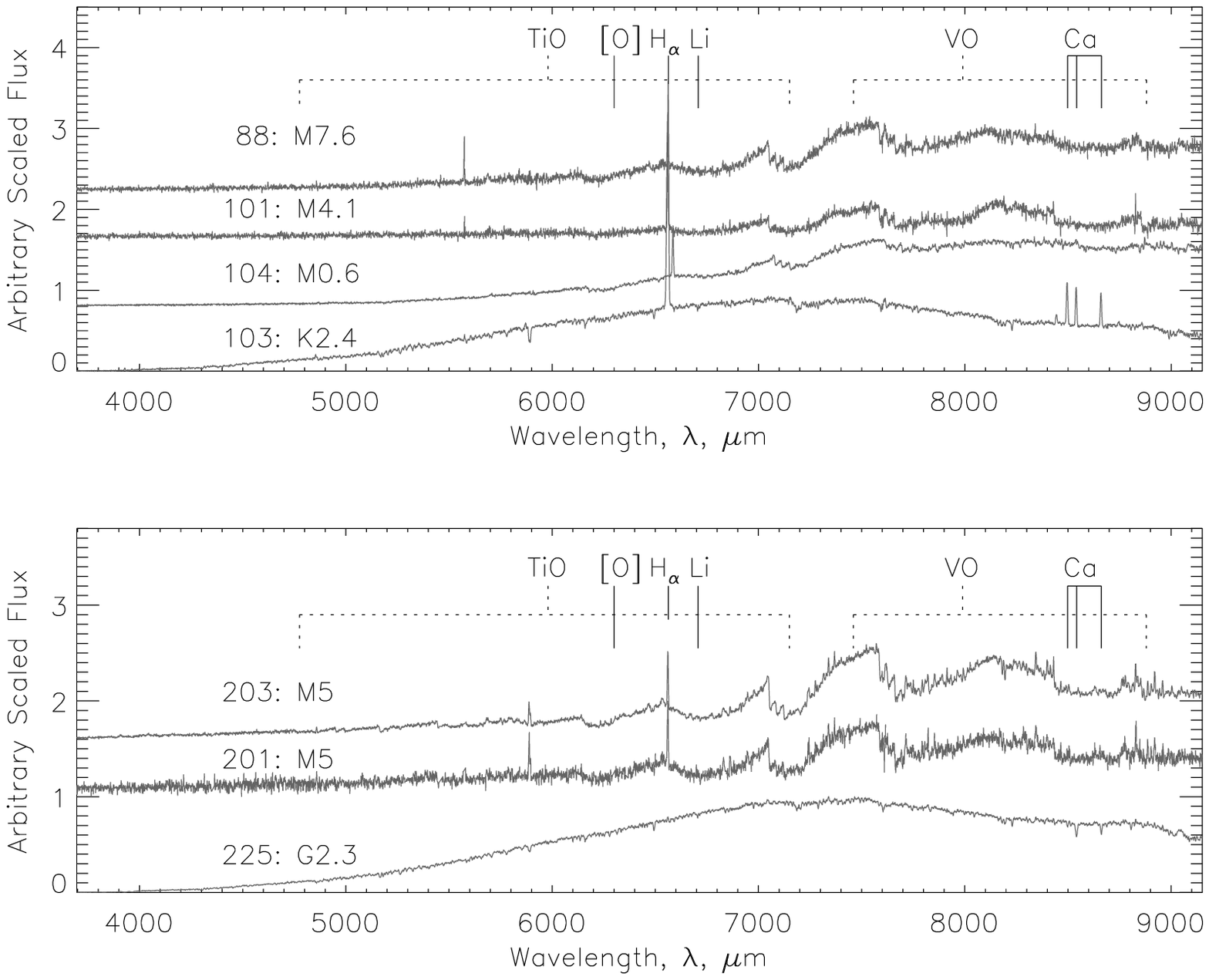}{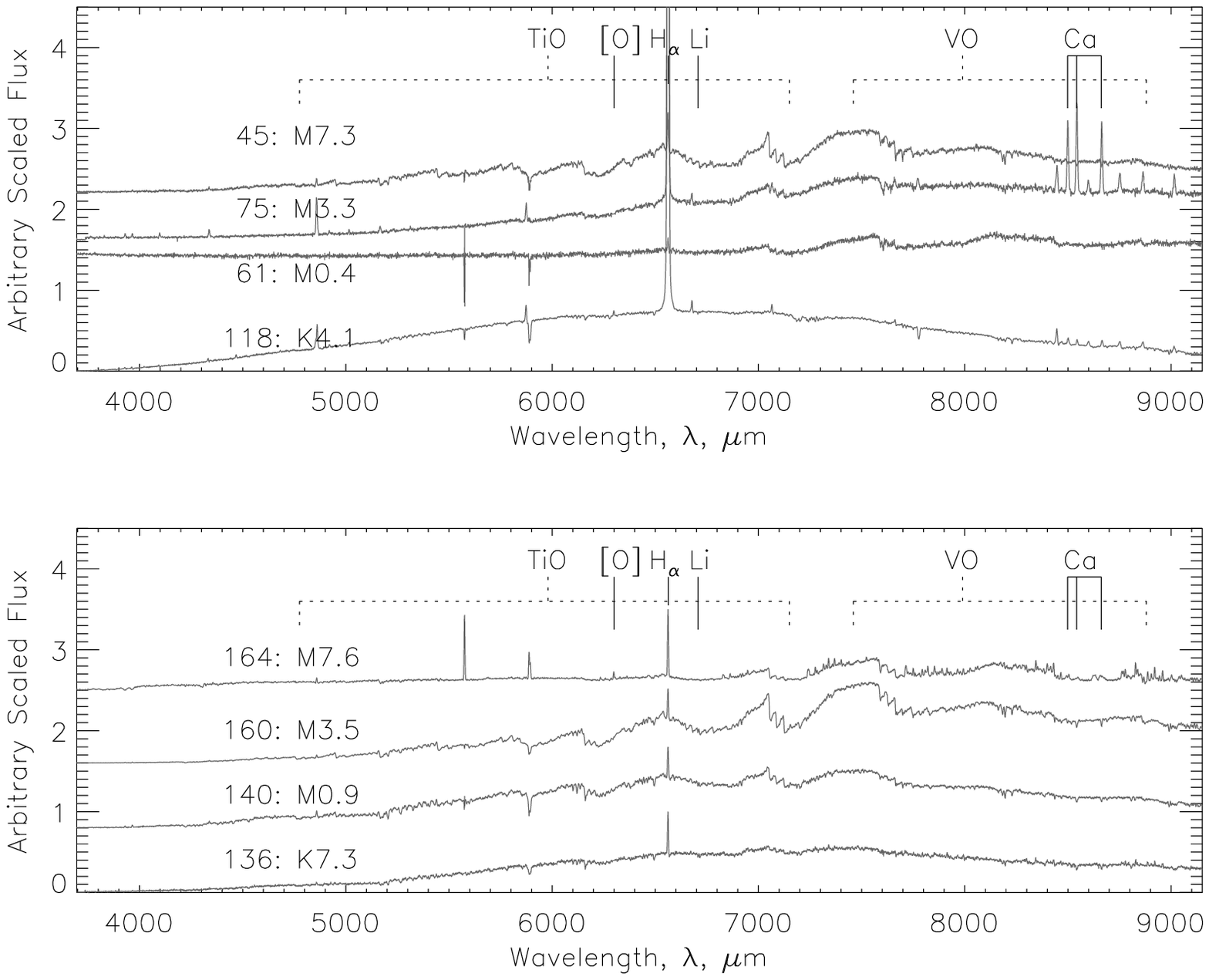}
\caption[Hectospec Spectral Sequence]{ 
A sequence of spectra in the Serpens cluster (left) and NGC 1333 (right) obtained with the Hectospec 
instrument.   The spectra are arranged in order from early type to late type stars, for a sample of Class II 
(upper plots) and Class III objects (lower plots).   
The TiO features used as a primary typing feature for late-type objects in the SPTCLASS code become 
progressively more prominent through the M class.
The strong feature at $6560\AA$ is the $H\alpha$ line of the Balmer series, an accretion signature in 
young stars.  In the sequences, the strong emission lines beyond $8000\AA$ are $Ca$, another 
strong indicator of accretion in young stars.   

}
\label{hectosern13}
\end{figure}

A total of 79 spectral classifications were obtained for YSOs observed with the Hectospec 
instrument, 32 in Serpens and 47 in NGC 1333.  
In addition, nine spectra from the randomly selected stars showed evidence of Lithium 
absorption, and are probably YSOs, c.f. Sec~\ref{lithium}.
The spectral classification of the Hectospec sources was carried out using  
{\it SPTCLASS: SPecTral CLASSificator code}
{\footnote{http://www.astro.lsa.umich.edu/$\sim$hernandj/SPTclass/sptclass.html}}, 
an equivalent width based method of 
spectral typing documented in  \citet{her2004,herSPT}. The code provides automatic spectral types 
and uncertainties to large samples of stars, these are then validated by eye.
Three different schemes are employed to type objects in three mass ranges of young stars: the 
late-type scheme for T-Tauri stars (K5 and later), the G-type scheme for 
intermediate mass T-Tauri stars  (F0 - K5), and the Herbig AeBe scheme for objects F5 
and earlier.  The majority of objects in the sample were late-type objects; 
the features utilized in the schemes are available in \citet{herSPT}. 
The equivalent widths of the $H{\alpha}$ and $Li~I$ features were measured using 
the {\it SPLOT} routine in {\it IRAF}  \citep[Image Reduction and Analysis Facility][]{tody}.   
In Fig.~\ref{hectosern13}, samples of the Hectospec spectra from the Serpens and NGC 1333 
clusters are shown, by evolutionary class, with their assigned spectral classifications. 
The TiO and VO features may be seen to become more pronounced in the later types. 
Table~\ref{hectotableS} lists the identifiers, spectral classifications, and $H{\alpha}$ and $Li~I$ 
equivalent widths  of the sources observed in Serpens.  Table~\ref{hectotableN} lists 
similar information for the NGC 1333 targets.
The effect of reddening on the Hectospec spectral classifications was examined to determine if high 
values of $A_{K}$ would create a bias to a later classification.  
The 300 spectra from each Hectospec observation were artificially reddened  by a further 
0.5 and 1.0 $A_{K}$ and rerun through the $SPTClass$ code.   In both cases, the spectral 
typing method was found to be stable, with the majority of the spectra returning the same 
classification.  A small number of the more noisy spectra were re-classed by no more than 
1.5 subclasses.  The signal-to-noise of these spectra was very low and they were not included 
in the final list of spectra.  
The SPTCLASS code uses equivalent widths which may be affected by veiling;   however we 
expect veiling will not significantly affect the spectral types as the method uses features over a 
wide spectral range. The program then determines a weighted mean spectral type, rejecting 
widely discrepant values.  Since it is unlikely that the veiling will bias the spectral type by the 
same amount for every diagnostic feature, the signature of veiling would be the lack of 
agreement between the different features.  We checked each source to ensure that the spectral 
types assessed from the individual features are consistent.

\subsubsection{SpeX}

\begin{figure}
\epsscale{1.}
\plotone{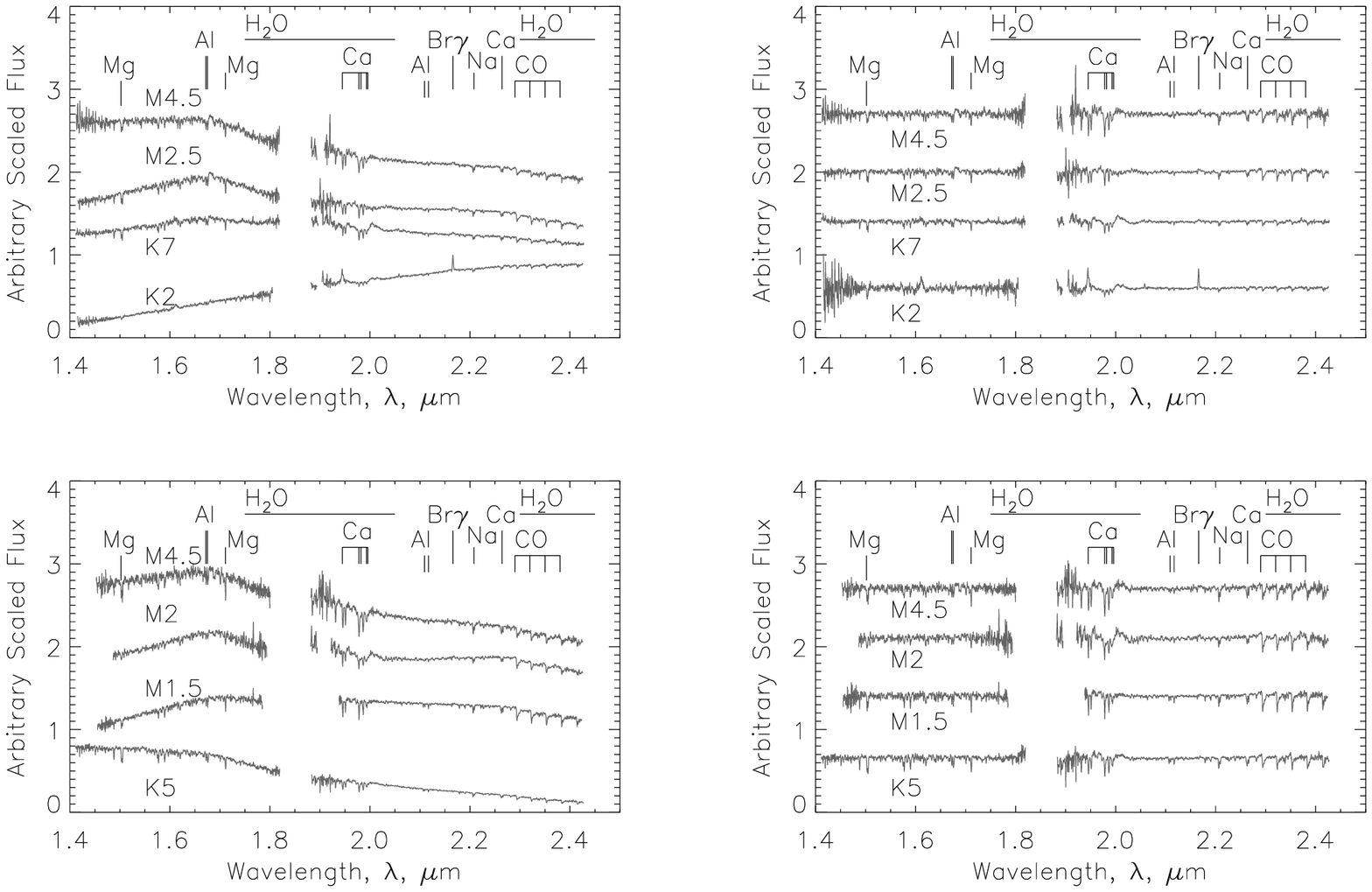}
\caption[SpeX Spectra: Serpens]{ 
Near-IR ($H$- and $K$-band) spectra taken with $SpeX$ of Serpens for a selection of Class II (above) 
and Class III (below) sources.   The left panel shows the raw spectra, the spectra on the right have 
been flattened by dividing out a highly smoothed version of the spectrum, all have arbitrary fluxes.  
}
\label{spextypingS}
\end{figure}

\begin{figure}
\epsscale{1.}
\plotone{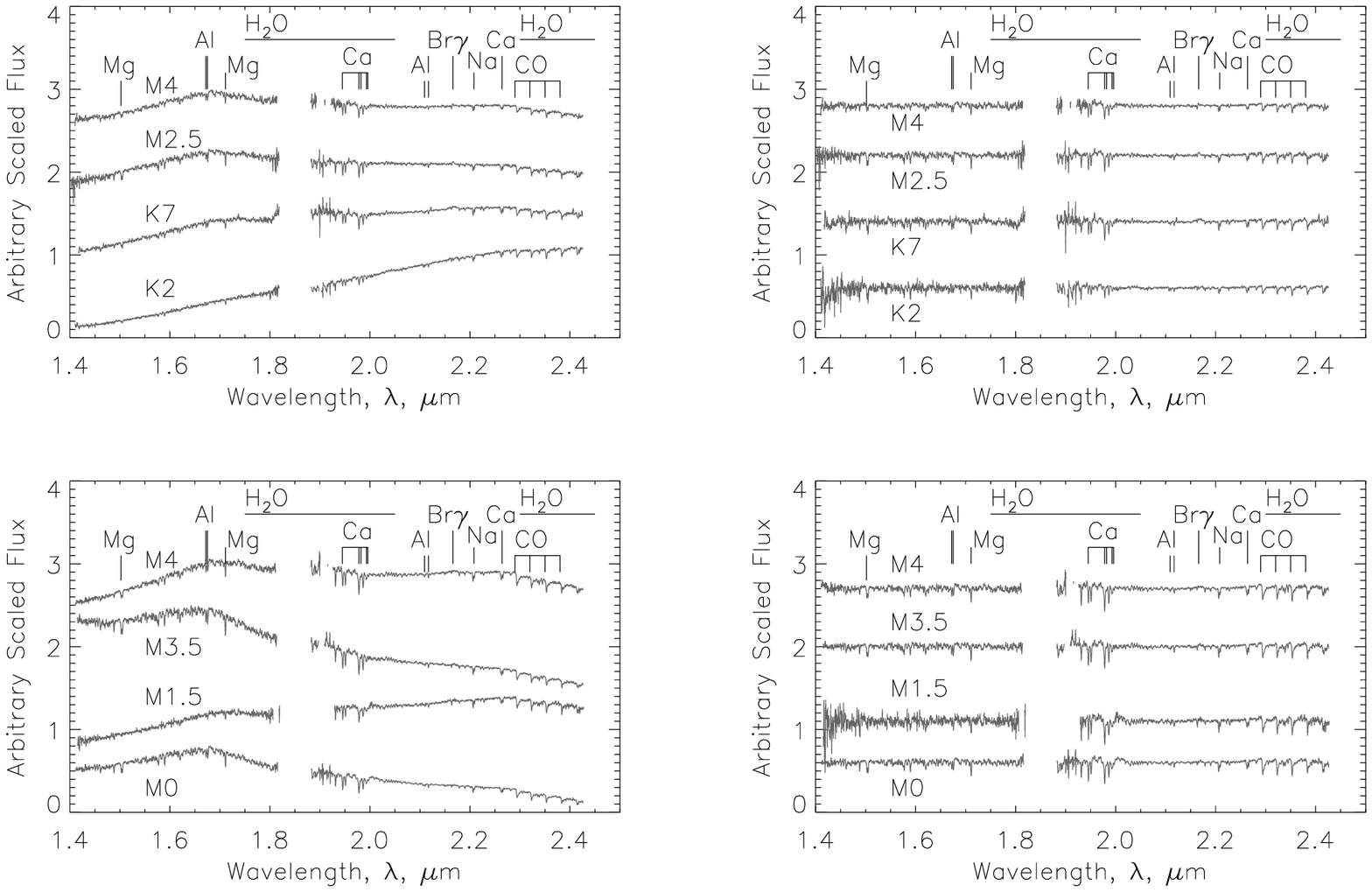}
\caption[SpeX Spectra: NGC 1333]{ 
Near-IR ($H$- and $K$-band) spectra taken with $SpeX$ of NGC 1333 for a selection of Class II (above) 
and Class III (below) sources.   The left panel shows the raw spectra, the spectra on the right have 
been flattened by dividing out a highly smoothed version of the spectrum, all have arbitrary fluxes.   
}
\label{spextypingN}
\end{figure}

We obtained spectra of 30 sources in the central region of the Serpens Cloud and of 22 sources 
in NGC 1333.  The spectral typing of the near-IR spectra was carried out by a visual comparison 
between the object spectra and standard spectra.  
The standards were obtained with the SpeX instrument as part of the IRTF Spectral Library and 
are publicly available online\footnote{http://irtfweb.ifa.hawaii.edu/$\sim$spex/WebLibrary/index.html} 
\citep{cus2,ray2}.  The spectral typing was carried out in two stages: initially, the cleaned spectra were 
examined and their continuum shapes, the $H_2O$ features, any emission features (such as $Pa{\beta}$ 
and $Br{\gamma}$) were noted and an initial spectral class was assigned to each.
The $J$-band was often dominated by noise, so further spectral typing was performed using the 
$H$ and $K$-bands only.  
Secondly, classification was carried out by comparison of the relative depths of specific 
spectral features to those of the library standards.  
An IDL routine produced a plot of the object spectrum from 1.45 to 2.35~${\mu}m$, from which 
the continuum component had been removed by dividing out a highly smoothed version of the 
spectrum, using the IDL routine $Smooth.pro$, with box size of 150. This routine allowed the 
overplotting of each of the standards in turn, to assess the fit to the object spectra. 
Figure~\ref{spextypingS} \& \ref{spextypingN} provides samples of the spectra obtained with SpeX, 
by evolutionary class, in each cluster.   The upper (lower) plots give examples of Class II (Class III) YSOs.  
The left and right-hand figures show the same spectra before and after continuum subtraction. 

The luminosity class and $v~sin(i)$ could be adjusted to obtain a best fit to the data.    
The surface gravities of young stars are generally lower than those of their older counterparts 
in the same spectral classification, having not yet contracted to their main sequence radii. 
It has been found by \citet{luh1} that averages of main sequence (V) and giant (III) standard 
spectra provide good approximations of the gravity-dependent features of young pre-main sequence 
stars.  The $CO$ lines that form a prominent feature from 2.3-2.4 ${\mu}m$ are gravity 
dependent.  The IRTF luminosity class III and V standards could be combined and overplotted, and were found 
to provide better fits, in many cases, to the $CO$-features particularly of the YSO spectra.
Of the observed features, the $Al$ features at 1.67 \& 2.11, the $Na$ feature at 2.2~${\mu}m$, 
and the $Mg$ feature at 1.71~${\mu}m$ were used primarily to ascertain the spectral classification 
\citep{cus2,ivan}.  The 2.11~${\mu}m$ $Al$ line is stronger than the 2.12 line in early types, 
transitioning near M0, and becoming weaker through the M-class, the $Al$ 1.67~${\mu}m$ feature 
varies similarly.  The relative depths of the lines in the $Na$ doublet at 2.21~${\mu}m$ and 
$Mg$ feature at 1.71~${\mu}m$ gradually grow more similar through the K and early M-class.     
The spectral types were determined by the relative depths of immediately adjacent features 
whose ratio should not be affected by veiling.  The lack of dependence with veiling was further 
demonstrated by using an IDL routine to apply veiling to the spectral standards.  Foe each object, 
veiling was applied to the standard spectra, but did not alter our determination of the spectral classification.  
By comparing each of the YSOs to standards with earlier and later spectral types than the best fit 
standard, we estimate the uncertainty for the SpeX spectral types to be 1.5 spectral subtypes.

The standard library is not complete for stars earlier than F, and so the classifications of the SpeX 
A-type stars were made using the calibration standard and from spectra in the literature \citep{wall,wallk,meyer}.  
The spectral types are therefore less certain and are listed without uncertainties. 
Table~\ref{spextableS} lists the identifiers and spectral classifications of the sources observed in Serpens.  
Table~\ref{spextableN} lists similar information for the NGC1333 targets.

\subsubsection[Consistency of Spectral Classifications]{Consistency of the Hectospec \& SpeX Spectral Classifications}\label{combsh}

\begin{figure}
\epsscale{1}
\plotone{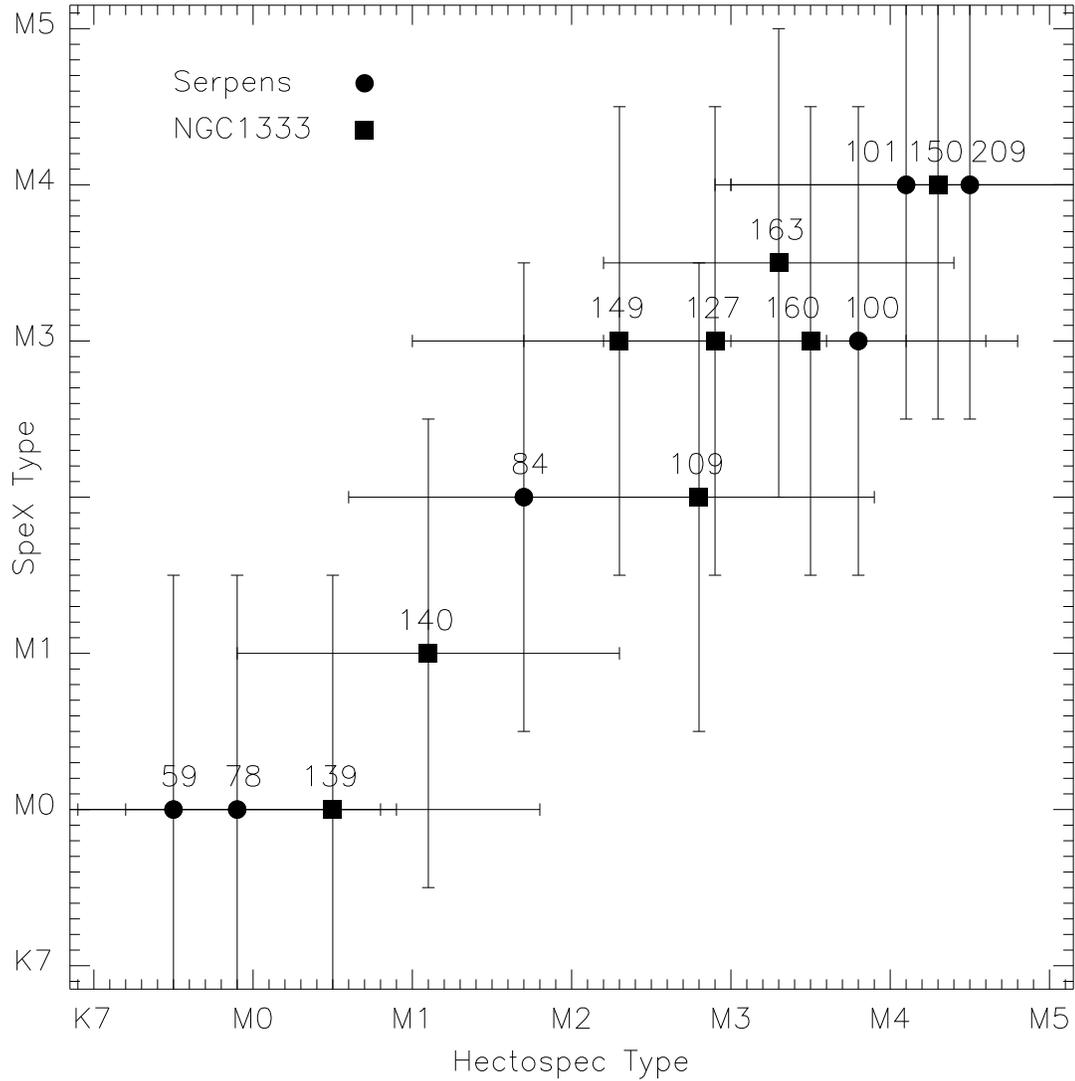}
\caption[Consistency of Spectral Classifications]{ 
The spectral classification for the YSOs with spectra taken with both Hectospec and SpeX were plotted to examine the 
consistency of the spectral typing methods in the far-red and near-IR.  The largest difference in classification was of 1 
spectral subclass, with some as small as 0.1 spectral subclass. The numbers refer to the source ID of the YSO, listed 
in Tables 1 - 4. 
}
\label{conspc}
\end{figure}

There were 6 stars in Serpens and 8 in NGC 1333 that were observed by both the Hectospec and SpeX instruments. 
Of the six objects in the Serpens cluster, [59, 78, 84, 100, 101, 209], five have spectral classifications 
from the two methods within 0.5 subclass of each other.  One of the objects, [100], lies within 1 subclass. 
In the NGC 1333 cluster, six of the objects, [127, 139, 140, 150, 160, 163], were spectral typed to within 0.5 spectral 
subclasses by the two methods. The remaining two objects, [149,109], were within 1 subclass. 
In each case, the difference in classification is well within the uncertainty listed for both methods.  
Where reddening was high, it was concluded that the SpeX spectral classification was the more accurate, and this 
was used in the subsequent analysis, for the less reddened sources, the Hectospec classification was used.   
Figure~\ref{conspc} shows the spectral classifications of the fourteen overlap objects, detailing the variation in spectral classification between the Hectospec and SpeX typing methods.

\subsection{Cluster Membership}

\subsubsection{Surface Gravity}

\begin{figure}
\epsscale{0.8}
\plotone{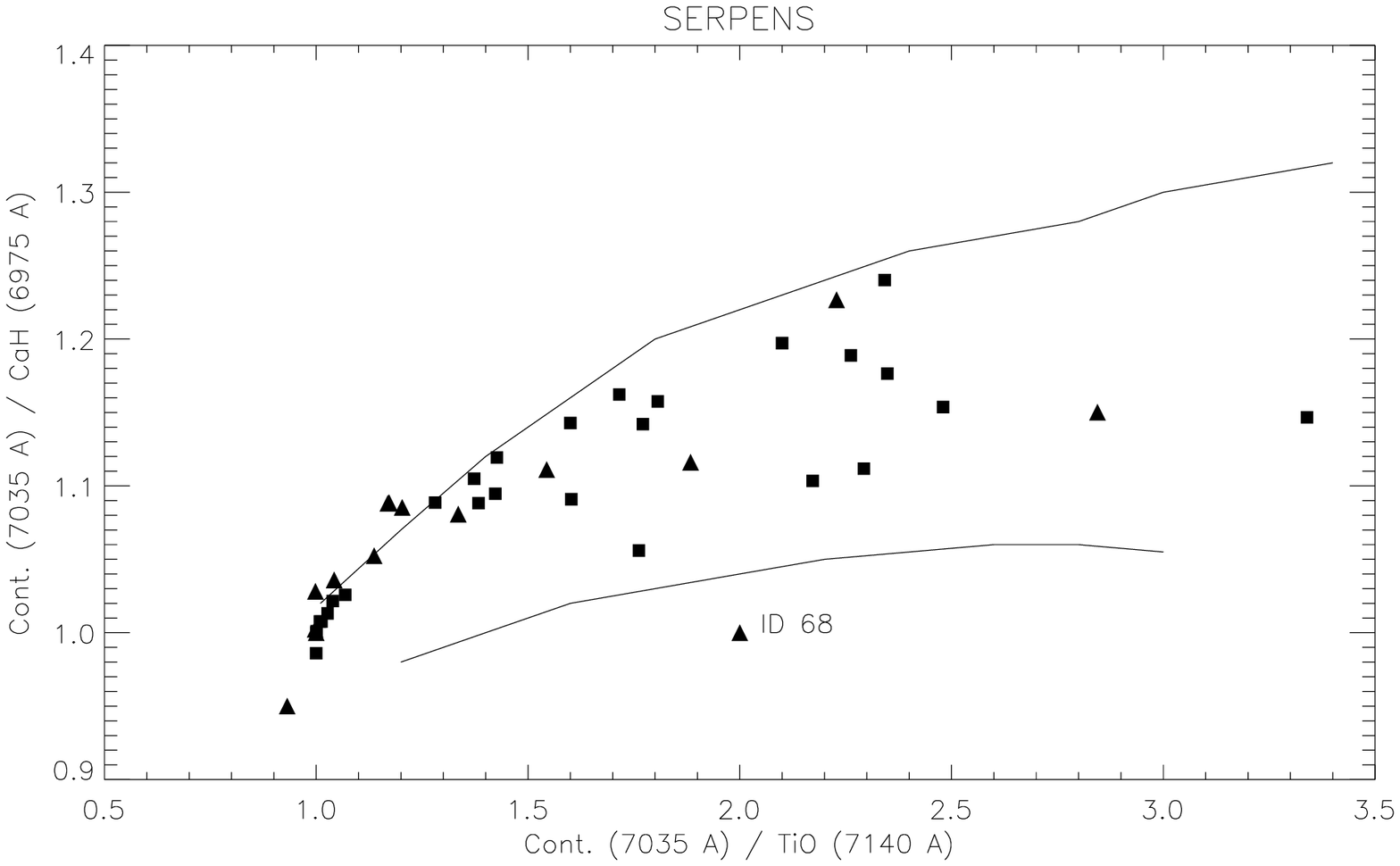}
\plotone{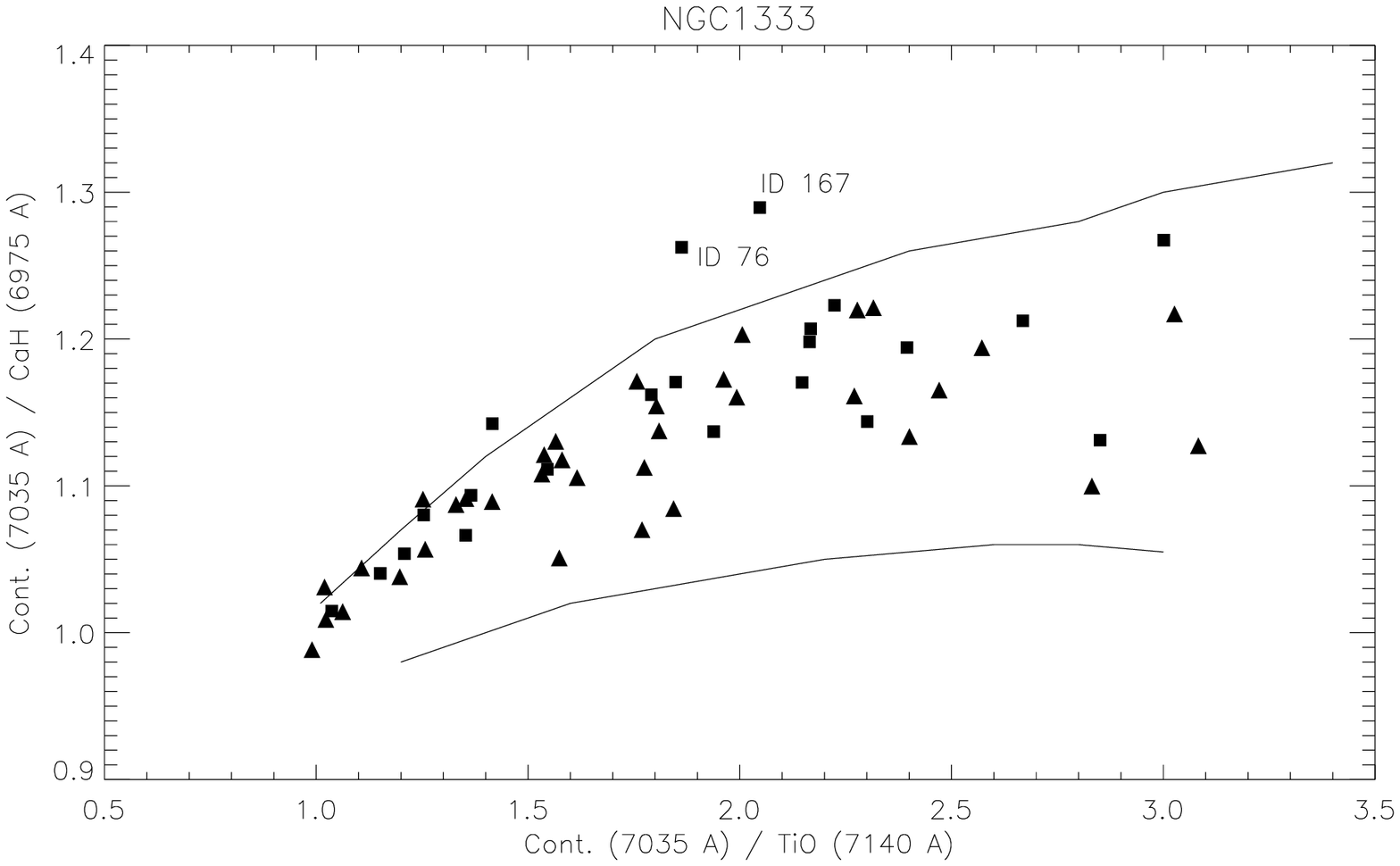}
\caption[Surface Gravities]{ 
The gravity sensitive features, $CaH$ and $TiO$ at 6975~$\AA$ and 7140~$\AA$, respectively, 
are plotted in ratio to the continuum at 7035~\AA.  The solid lines in each plot represent the 
locii of the dwarfs (upper) and giant (lower) stars.  In both clusters, Serpens (above) and 
NGC 1333 (below), the objects fall mainly in the intermediate region for surface gravity, 
the domain of the young stellar objects.   The sources with ages $>$3~Myrs are shown by 
squares, those with ages $<$3~Myrs by triangles.   
One source in Serpens appears to be a giant [ID 68].   In NGC 1333, there is some scatter 
into the dwarf regime [ID 76, 167].   Due to the uncertainties in this measurement, we still 
consider these candidate YSOs and have not rejected them from the sample.
}
\label{surfgrav}
\end{figure}

The surface gravities of young stellar objects lie in an intermediate regime between 
dwarf and giant stars, for a given spectral type.  The determination of the approximate 
surface gravities for stars in our sample is performed as a check on the likelihood of 
their being young stars and not field dwarf or background giant contaminants in the sample.    
Following \citet{all}, the strengths of the gravity sensitive $CaH$ and $TiO$ 
features centered at $6975~\AA$ and $7140~\AA$, respectively, were measured.  
Given that these features lie in the far-red, only those sources with Hectospec spectra 
could be plotted.  The ratio of each feature over a $\pm15~\AA$ band was taken to the 
continuum at $7035\pm15~\AA$.  Figure~\ref{surfgrav} shows the $CaH$ to $TiO$ ratios 
for Serpens (above) and NGC 1333 (below).  
The upper solid line indicates the loci of dwarf stars, the lower line that of the giants.  
In both clusters, the spectral sources are seen to fall into the intermediate 
region as expected of YSOs still contracting to their Main Sequence radii.  
Three objects (ID 76, 167 in NGC 133, ID 68 in Serpens) lie outside the expected region, 
though due to the low resolution (and thus high uncertainties) we do not reject them as YSOs. 
We have also used the ages determined from the HR diagram (c.f. Section~\ref{hrds}) to examine 
whether or not those objects with ages $>$3~Myrs (shown as squares in Fig.~\ref{surfgrav}) are 
more likely to have lower surface gravities (similar to dwarf stars) than those $<$3~Myrs (the 
triangles).  
No obvious trend in surface gravity with age is evident  in the data; this is strong evidence 
that the older stars are not contamination from background giants with dusty envelopes or 
foreground dwarfs with detectable X-ray emission.   There is some indication in NGC 1333 
that the older stars have more  dwarf-like surface gravities,  which if verified, could further 
substantiate their older ages.   A more precise surface gravity determination will require 
high-resolution echelle spectra of the  clusters.

\subsubsection{Lithium}\label{lithium}

The $Li~I$ doublet, at 6707~$\AA$, is well known as an indicator of stellar youth, and  
was observed in the Hectospec far-red spectra, where its equivalent width (EW) was 
measured for all typed spectra.  Lithium is depleted rapidly once a star begins to 
undergo nuclear fusion reactions in its core, and so its presence in absorption is 
a good indicator of the pre-main sequence nature of a star and can be used to 
verify membership.  If, however, the star is not massive enough to undergo fusion, 
then no depletion occurs, so this method is not reliable for substellar objects.    
More massive objects develop radiative zones that prevent convection of $Li$ 
to the core, and so depletion occurs more slowly. The most rapid depletion occurs 
in the M0-M5 range.     An equivalent width (EW) between 0.1 and 0.8~$\AA$  can be 
considered as a valid indicator for the presence of Lithium in absorption \citep{bas,mar1,mar2,bri}.   
As the Hectospec instrument provides up to 300  spectra in each observation, many stars 
which met our magnitude and color criteria were randomly selected in the Serpens and 
NGC 1333 fields  to search for Class III members without detectable X-ray emission.
These sources may still show detectable $Li~I$ absorption in their spectra, indicating 
that they are young stars, and not evolved field stars in the line of sight.  
Any  previously unidentified object 
with a good spectral detection and evidence of $Li~I$-absorption was flagged and its 
reddening checked for consistency with known cluster objects.  
Given the moderate resolution of the Hectospec data ($R \approx 3000$), and the 
presence of lines of Fe I, V I, CN and TiO in the vicinity of the $Li~I$ feature \citep{bas}, 
it is not possible to unequivocally state that these objects are members.  
Four objects in Serpens and five objects in NGC 1333 were identified as possible 
cluster members by this method (c.f. Table~\ref{hectotableS} \& \ref{hectotableN}).  
As they exhibit no IR-excess they are in the evolutionary Class III stage.  
Seven of the nine have spectral types M1-M4, the remaining two have spectral types 
K5 and K6.5.  Assuming rapid $Li~I$ depletion for the M stars, and some depletion for the K stars, 
detection of $Li~I$ would constrain their ages to be $<$3~Myrs to $\sim$10~Myrs. 
Their ages, as determined from the HR diagram (c.f. Section~\ref{hrds}), give ages for the 
K stars and two of the M stars 10-20~Myrs, while the remaining five M stars have ages 2-5~Myrs.
Twelve of the previously identified Serpens YSOs, and thirty-two of the NGC 1333 YSOs 
show evidence for Lithium in absorption in their spectra.

\subsubsection{$H{\alpha}$}

\begin{figure}
\epsscale{0.8}
\plotone{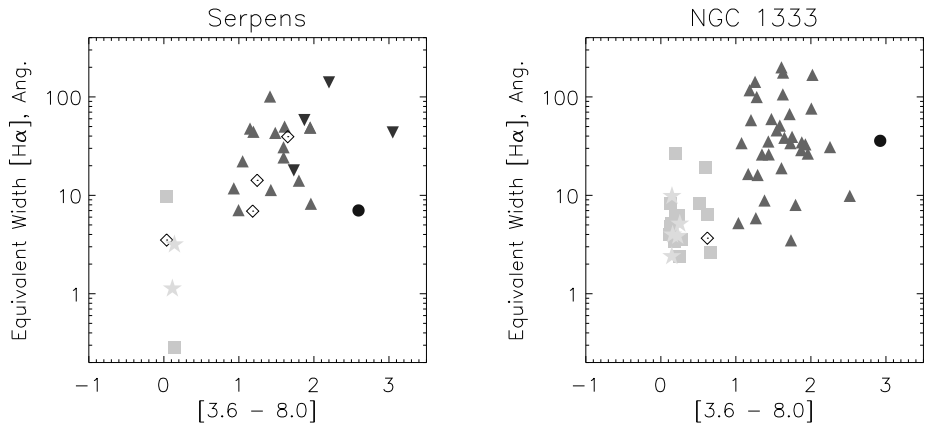}
\caption[$H{\alpha}$  with IR-excess color]{ 
The equivalent widths of the $H{\alpha}$  lines with IR-excess color for Serpens and 
NGC 1333.  Symbols indicate Class I (circle), flat spectrum (inverted triangle), 
Class II (triangle), transition disk (diamond), Class III (square), $Li~I$ candidate Class III (stars).   
We find the sources with significant [3.6]-[8.0] excesses show substantially higher 
equivalent widths, as would be expected if the $H{\alpha}$ emission is created by 
accretion from a circumstellar disk.
}
\label{hlc}
\end{figure}

\begin{figure}
\epsscale{0.9}
\plotone{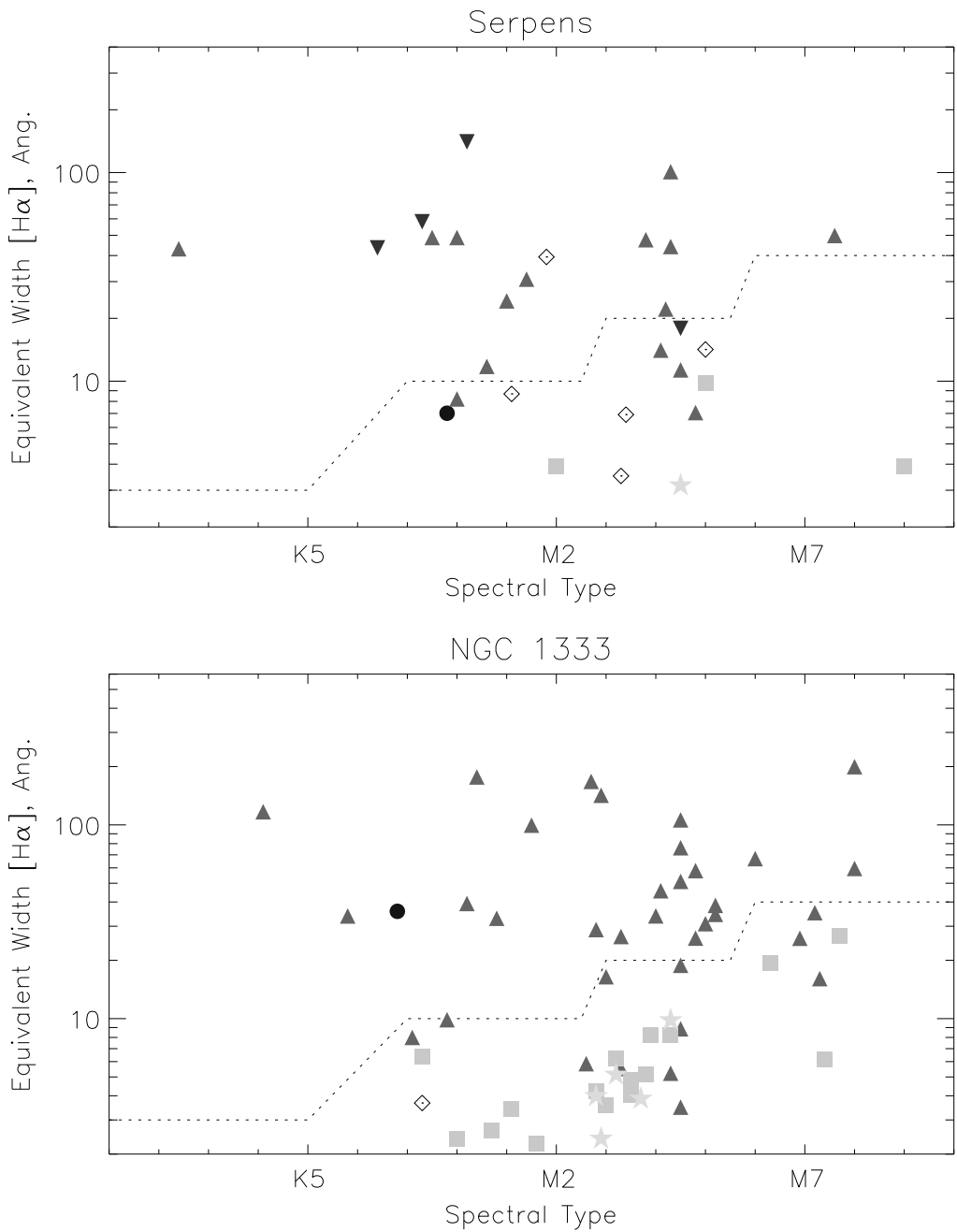}
\caption[$H{\alpha}$ with Spectral Classification]{ 
The equivalent widths of the $H{\alpha}$ line with spectral classification for Serpens and NGC 1333.  
Symbols indicate Class I (circle), flat spectrum (inverted triangle), 
Class II (triangle), transition disk (diamond), Class III (square), $Li~I$ candidate Class III (stars).   
The dotted lines indicate the level at which $H{\alpha}$ emission is thought to no longer arise purely from 
chromospheric emission but also from accretion onto the star \citep{whi}.   
None of the Class III objects appear to be accreting, while most of the Class II objects are accreting.    
}
\label{hls}
\end{figure}

Another commonly studied spectral feature in young stars is the $H{\alpha}$ line at 
6563~$\AA$.  Emission in the $H{\alpha}$ line can arise from chromospheric activity, 
but it also traces accretion onto the surface of a star from a disk.  
In Fig.~\ref{hlc}, the $H{\alpha}$ EW  is plotted against the IR-excess color of 
the star to compare the accretion properties with the evolutionary class of the source. 
Objects with a high $3.6-8.0$ color, (the disk-bearing stars), show higher levels of $H{\alpha}$, 
usually above 10~$\AA$, than those with less (the diskless stars).       
Figure~\ref{hls} shows the equivalent width of $H{\alpha}$  with respect to the spectral 
type of the star.  All chromospherically active stars will emit some level of $H{\alpha}$ 
which increases with later spectral classification; the dashed line on the $H{\alpha}$ plot 
indicates the level of emission required to indicate that accretion is occurring \citep{whi}.  
It can be seen that the vast majority of Class II sources are still accreting from their disks.
Even sources which fall below the Ha EW threshold may be accreting;
\citet{flamuz} recently found sources with low $H{\alpha}$ EW show evidence of 
accretion in higher resolution spectroscopy.
In contrast, the Class III YSOs do not show emission above the level expected from chromospheric activity alone.  
The transition disks, in all but one case, show no evidence that they are accreting 
across their inner disk gap.  None of the young cluster members identified by the $Li~I$ 
line in their spectra show evidence of accretion, in keeping with their showing no evidence 
of optically thick circumstellar material. 
Eleven previously identified YSOs in Serpens and twenty in NGC 1333 show $H{\alpha}$
EW indicative of accretion.

\subsubsection{Other Emission Lines}
Four other emission lines of interest were observed: the $Ca II$ triplet at 8498, 8542, and 
8662~\AA,  the $O I$ line at 6300~\AA,  the Hydrogen Paschen line, 
$Pa{\beta}$ at 1.282~${\mu}m$, and the Hydrogen Brackett line, $Br{\gamma}$ at 2.166~${\mu}m$.   
The $O I$ line is due to stellar/disk winds and outflows, while the other three arise 
from accretion processes, and indicate ongoing accretion from a circumstellar disk.   
Emission in the $Ca II$ triplet feature was observed in three of the four flat spectrum sources, two 
out of seventeen Class II members, and one of the five transition disks in Serpens. 
The $Ca II$ feature was also observed in the one Class I object and ten of the thirty-two Class II 
objects in NGC 1333.  All of these objects also have emission in the $H{\alpha}$ line.  
The $Br{\gamma}$ line was observed in emission in 3 of the 12 Class II Serpens YSOs observed with SpeX, 
with weak emission in a fourth, and in six of the thirteen NGC 1333 Class IIs.      
The four Serpens YSOs also had emission in the $Pa{\beta}$ line, while five of the NGC 1333 
Class IIs show emission in this line.   
Emission in the $O I$ feature was observed in 13 sources in Serpens: the lone Class I, 2 of the 4 
flat spectrum sources, 6 of 17 Class IIs, 2 of 5 transition disks, and 2 of the 4 candidate Class IIIs.  
In NGC 1333, fifteen YSOs were observed to show $O I$ in emission: again, the lone Class I, 12 of the 
32 Class II sources, and 2 of the 13 Class III stars.

\subsubsection{Summary of Membership}

In total, spectral types were obtained for 65 members in the Serpens clusters, 30 with SpeX 
and 35 with Hectospec. Of these,  35 had detections in the {\it Chandra} X-ray data. 
Tables~\ref{hectotableS} \& \ref{spextableS} list the sources for which spectra were 
obtained in the Serpens cluster, along with spectral classification, calculated bolometric 
luminosity, equivalent widths of the $H{\alpha}$ and $Li~I$ lines, and whether any other 
emission line was detected.    
The photometry of the candidate Class III members in Serpens identified solely through 
the detection of $Li~I$ absorption in their spectra is given in Table~\ref{tblsernewcIII}.
In the NGC 1333 cluster, spectral types were obtained for a total of 74 YSO, 22 with SpeX 
and 52 with Hectospec,  55 of which had detections in the X-ray data. 
Tables~\ref{hectotableN} \& \ref{spextableN} list the objects in the NGC 1333 cluster 
for which spectra were obtained.    
The photometry of the new candidate Class III members identified in NGC 1333 by $Li~I$ 
absorption in their spectra are given in Table~\ref{tbln13newcIII}.

\begin{figure}
\epsscale{0.8}
\plotone{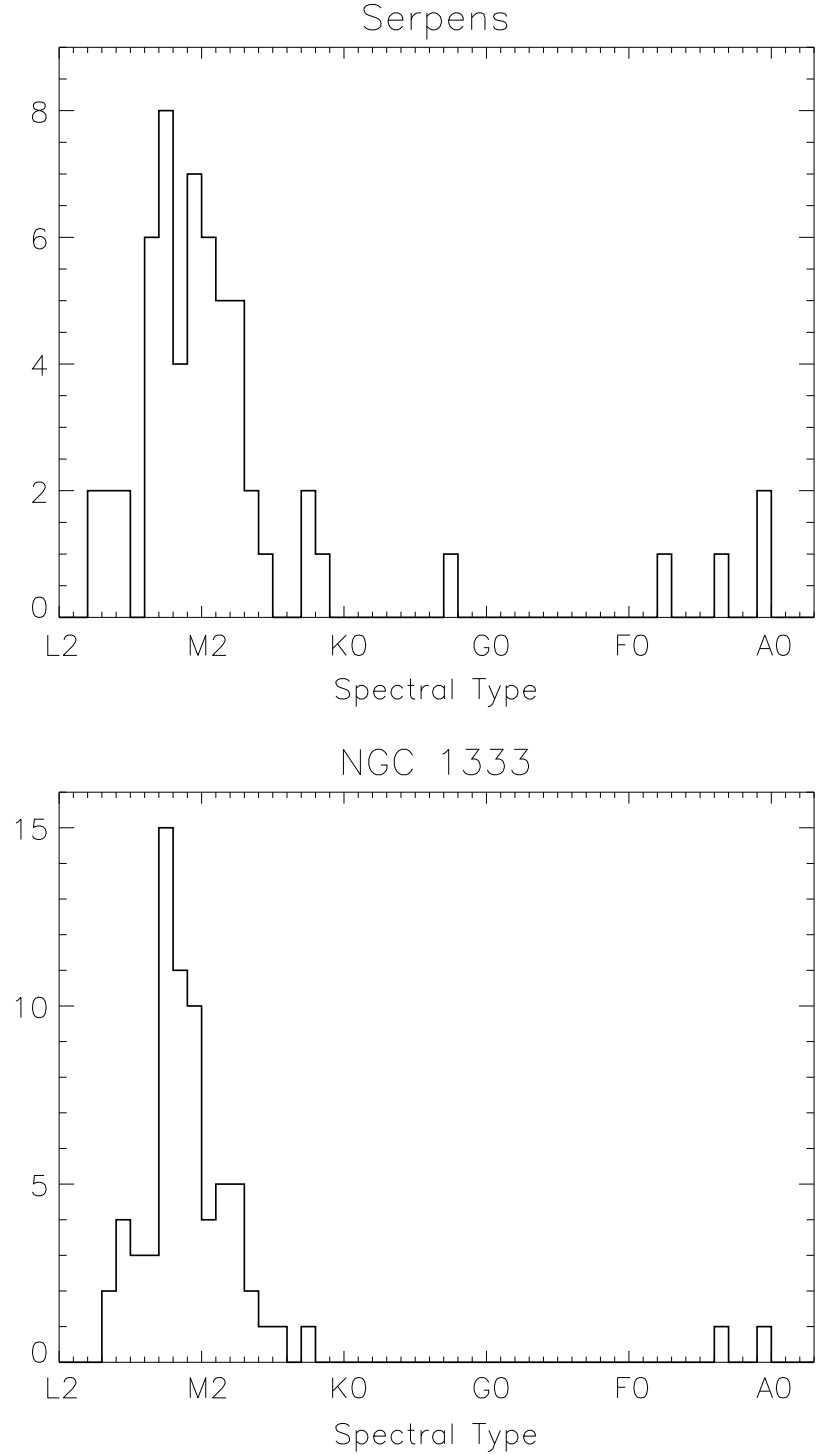}
\caption[Histogram of Spectral Classifications]{ 
Histograms of spectral classification for the 64 objects in Serpens and 69 objects in NGC 1333, for which 
spectral classifications were obtained for known cluster members.  
The clusters are both low mass, with the majority of objects either M or K spectral classification.    
}
\label{histo_types}
\end{figure}

The evolutionary classifications of these spectral sources are summarised in Table~\ref{goodspecclass}. 
The table lists the numbers of spectra obtained in each cluster by class and X-ray 
detection, for each of the clusters. The classes are grouped as Class I (I), flat spectrum (FS), 
Class II (II), transition disk (TD), Class III (III), and those Class III objects that 
were not detected in X-rays but exhibited a possible $Li~I$ feature in their spectra ([III]).
The majority of the spectral classifications obtained were Class II or Class III, the more deeply 
embedded protostars are more difficult to accurately type as the scattered light from  
their envelopes and emission from their accretion disks can alter the features of their spectra. 
Figure~\ref{histo_types} shows the histograms of spectral classifications for known members of 
the two clusters; both clusters consist primarily of low mass stars with  K and M spectral types, 
while the highest mass stars classified in each cluster are A-type.  
In total, the membership of  27 Serpens YSOs and 48 NGC 1333 YSOs were independently 
confirmed spectroscopically by $Li~I$ absorption or accretion signatures.  
Further verification of the pre-main sequence status is given for the 85 sources with far-red spectra 
by the surface gravity indicators;  only one possible background giant in Serpens and two possible 
foreground dwarfs in NGC 1333 were identified although there are large uncertainties in the 
surface gravity indicators.   
The remaining 28 Serpens YSOs and 15 NGC 1333 YSOs could not be independently confirmed 
by our spectroscopic observations; however, the presence of X-ray emission and/or infrared-excess 
makes all of these likely members.  As we will discuss in the next section, there is a small 
incidence of contamination by background AGB stars.

A comparison of the NGC 1333 spectra was undertaken with those of \citet{asp1,asp2,asp3}. 
A total of 24 of our sources with spectral types matched to sources in \citet{asp1,asp2}, with 
spectral types given by \citet{asp1} for 18 of the 24 sources.   Further, near-IR spectral types 
from \citet{asp3} also matched to 18 of the 24 sources.  
Where the later \citet{asp3} types were available, these were used in our comparison; 
ten of the sources have spectral types within 5 subclasses, eight sources move from early/late 
K to early/late M-class.   In the remaining five cases, the spectral types disagree by more than 
one spectral class.  The Aspin spectra cover only the $K$-band and have lower spectral 
resolution than those of this work.  The spectral classifications are roughly consistent, with 
K-M stars corresponding to K-M stars.

\subsection{HR Diagrams}\label{hrds}

Models of stellar  evolution \citep[see for example:][]{bar,dan} can provide an estimate 
of the age and mass of young stellar objects from their placement on the HR diagram. 
The models chosen for this study were those of \citet{bar}.   These tracks have been 
found to model stars M5 and later particularly well, where other tracks can assign 
systematically  older ages to sources with decreasing temperature \citep{luh1}.   The 
bolometric luminosities of the spectral sources were calculated from their magnitudes 
at $H$-band as follows:  
the bolometric correction is calculated for the $H$-band using the correction for the 
$V$-band and the color  term $[V - H]$.  The bolometric magnitude is then calculated 
using the $H$-band magnitude $m_H$, after correcting for the extinction present due 
to the ISM and circumstellar dust.   The extinction at $H$-band, $A_{H}$, is determined 
using the 2MASS and IRAC 4.5~${\mu}m$ colors and the \citet{fla} reddening law, as 
described in \citet{gutthesis}.   The bolometric corrections, $[V-H]$ colors, and effective 
temperatures at each spectral type  were taken from \citet{ken95} Table 5, with 
adjustments at later spectral classifications (M1 onwards) from \citet{luh1}.  These 
were used to determine the bolometric correction for the $H$-band and the bolometric 
luminosity in the following manner:
\begin{eqnarray*}
BC_{H} = BC_{V} + [V - H] \\
m_{bol} = m_H + BC_{H} - A_{H}   \\
M_{bol} = m_{bol} -5 \times log(D(pc)) + 5   \\
log(L_{bol}/L_{\sun}) = (M_{bol} - 4.76)/(-2.5)    
\end{eqnarray*}
The absolute bolometric magnitude and luminosity are then calculated assuming 
distances of 260 and 240~pc to Serpens and NGC 1333, respectively. The 
luminosities are given in  $log{L_{\*}}/{L_{\sun}}$.   Figures~\ref{hrdscx} and 
\ref{hrdncx} show the HR-diagrams of the young stellar objects with spectral types 
in Serpens and NGC 1333,  plotted according to evolutionary class and X-ray 
detection, respectively.   Adopting a typical uncertainty in the spectral types of 1.5 
subclasses, we calculate representative formal uncertainties of  0.2 in 
$log({L_{\*}}/{L_{\sun}})$  and 0.02 in $log(T_{eff})$, as shown on the HR diagrams.

In addition to the random uncertainties, a number of systematic uncertainties affect the 
determination of the temperature and luminosity of pre-main sequence stars.    
First, surface gravity can affect the depth of spectral features.  
\citet{hil} discuss the conversion from spectral classification to effective temperature 
derived for MS stars and find that they may not be applicable to their pre-main sequence counterparts.  
\citet{coh} calculate an overestimate in $T_{eff}$, for K \& M stars, of 0.01-0.03 in 
$log(T_{eff})$ or $\sim$1.5 subclasses. 
Secondly, the calculated luminosity has errors caused by absorption and scattering by 
circumstellar material (an inclination dependent effect), uncertainties in the reddening 
law and bolometric corrections, an uncertain contribution due to accretion luminosity and a 
variable stellar luminosity. 
\citet{ken2} predict a scatter in luminosities for Class II objects with the same age and mass of 
up to 1.0 in $log(L_{bol}/L_{\sun})$ due to the inclination of a circumstellar disk 
and accretion onto the stellar surface,  while \citet{cho} estimate an error of $\pm$~0.12  
in $log(L_{bol}/L_{\sun})$ due to variability.

Through a comparison of weak line and classical T-Tauri stars, \citet{cieza} showed the presence of 
excess emission in the $J$ and $H$-bands \citep[see also:][]{edw}.  Using the statistics in 
Table~6 of \citet{cieza}, we crudely assess the impact of this excess on our data.  Excess emission 
in the $J$ and $H$-bands not only increases the value of $m_H$ over the photospheric, but it also 
changes the estimate of the extinction.  The estimate of the extinction is driven primarily by the
$[J-H]$ color, and we can approximate the increase in the measured $m_H$ over the photospheric
value, $m_H(photo)$,  as
\begin{displaymath}
m_H = m_H(photo) + m_H(excess) - [J-H]_{excess} \times A_H/E(J-H)
\end{displaymath}
Using median values of $m_H(excess)$ = -0.47~mag, $m_J(excess)$ = -0.26 mag, and $A_H/E(J-H)$ =
1.66 \citep{fla},  we find that the increase in $m_H$ over $m_H(photo)$ is -0.8~mag.  This would result 
in an overestimate of $log({L_{\*}}/{L_{\sun}})$ of 0.3, or 1.5 times the typical uncertainty.  This 
overestimate in the luminosity will lead to an underestimate in age.   With the current data, 
it is impossible to determine the amount of H-band excess for our individual sources.  We note that
the excesses will only affect Class II objects and will have no affect on the Class III objects which do 
not have disks or accretion.   Thus, the good agreement in the ages of the Class III and Class II  objects 
shown in Sec.~\ref{disc} suggests that  the excesses in the $J$ and $H$-band do not strongly affect our 
luminosity, and hence age, determinations.

The effects of binarity on luminosity calculations are uncertain and model-dependent.  
The frequency of binarity in a cluster may be determined by the density 
of the cluster \citep{bou}, with a possible decline in frequency as a function of age 
\citep{ghe}.  The presence of a later-type companion would lead to an overestimate in 
$log(L_{bol}/L_{\sun})$ of up to $\sim$0.3.  
Uncertainty in the distances will lead to a shift in the calculated luminosities of the 
cluster members.  
If the distance to the cluster is overestimated/underestimated by a factor of 10\%, 
then the change in luminosity  will be $\pm 0.17$~$log(L_{bol}/L_{\sun})$.

Possible sources of contamination among the stars identified as cluster members include 
Asymptotic Giant Branch (AGB) stars. These objects are surrounded by dusty envelopes 
which produce infrared excesses in  the IRAC \& MIPS bandpasses \citep{rob}.  They are more 
luminous than low mass YSOs, and tend to lie above the calculated stellar birth-line. Six 
of the Serpens sources, [71, 122, 127, 134, 158, 170], are found to be possible AGB 
contaminants, while none are identified in NGC 1333. One of these has an optical spectra; 
the resulting surface gravity is consistent with a YSO; the remaining objects were observed 
only with SpeX, and we do not have information on surface gravity.  
Five of the sources have spectral energy distributions consistent with transition disks, with 
an excess at 24 ${\mu}m$ only.  However, these differ from transition disks in that they have 
[8]-[24] $<$ 2, while most transition disk systems have [8]-[24] $>$ 2  \citep{muz,flamuz}.

As a comparison with the \citet{bar} tracks, from which the ages and masses of the spectral 
sources are calculated, the HRDs were replotted using the models of \citet{dan} and the ages 
recalculated. These tracks tend to assign younger ages to stars earlier than M5, and older 
ages to those later than M5.  In NGC 1333, the median age for sources earlier and later than 
M5 for \citet{dan} was 1.0~Myrs and 3.5~Myrs, and for \citet{bar}, 1.7~Myrs and 2.4~Myrs.  
In Serpens, the median ages were: 2.6~Myrs and 3.1~Myrs, and, 2.6~Myrs and 4.6~Myrs, respectively. 
Comparisons between these two models gives some indication of the uncertainties still involved in 
the understanding and modeling of the physics behind the evolution of pre-main sequence stars, 
with issues such as the incompleteness in the opacity tables at low mass temperatures \citep{bar,lyra}. 
To minimize the uncertainties due to the pre-main sequence tracks, in the next section we rely on  
the relative ages derived with a single set of pre-main sequence tracks.

\section{\bf Discussion}\label{disc}

\begin{figure}
\epsscale{0.8}
\plotone{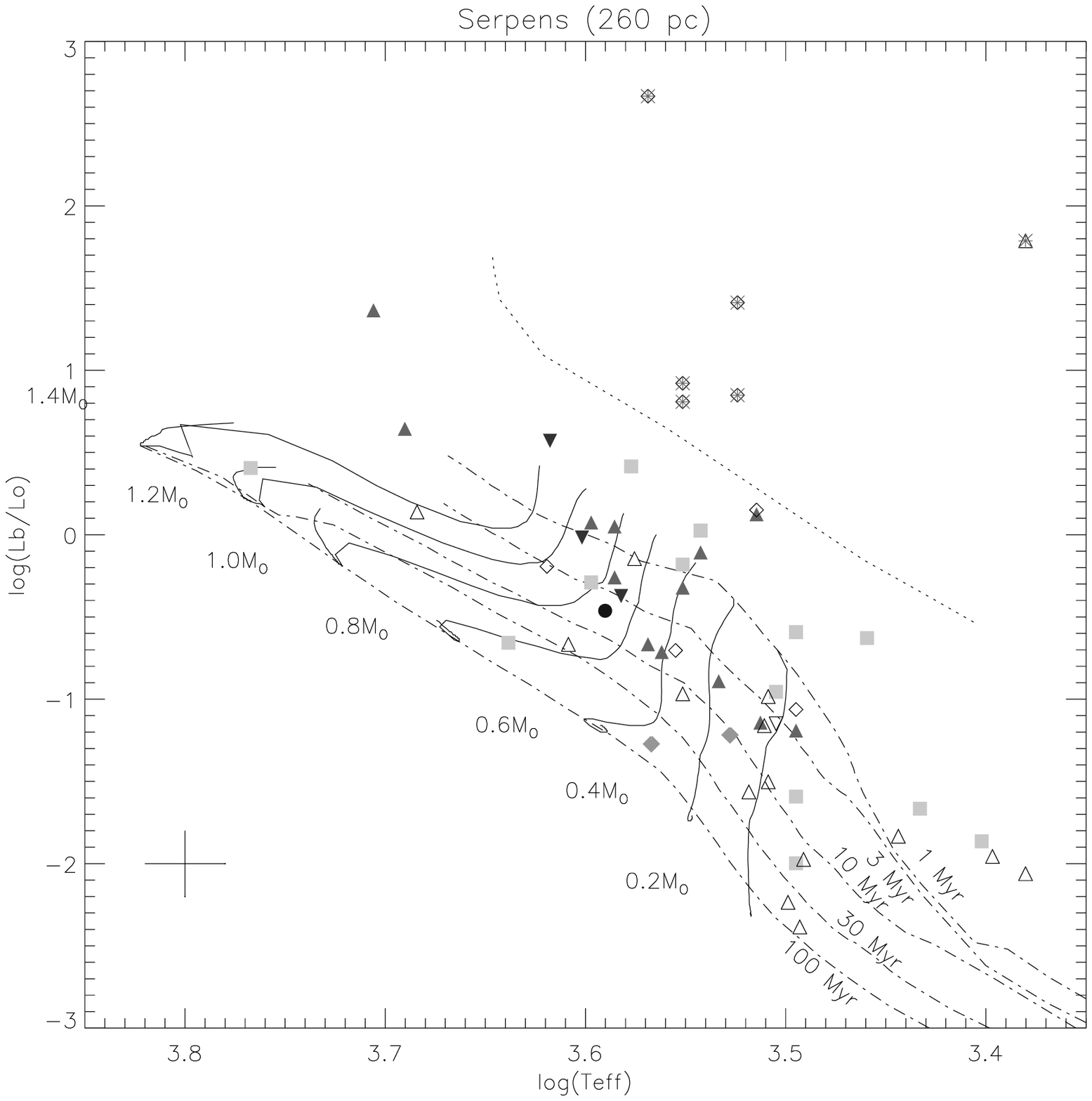}
\caption[HR diagram: Serpens by class \& X-ray detection]{ 
The H-R diagram of the Serpens cluster plotting all good spectral observations.  
The sources are identified by their evolutionary classification, with solid circles indicating 
the Class I protostars, the inverted triangles the flat spectrum.   The Class II objects are 
represented by triangles, transition disks by open diamonds, and Class III sources 
by squares.    The evolutionary tracks are taken from \citet{bar}, with the 
1, 3, 10, 30, and 100~Myr isochrones, and mass tracks from 0.2 to 1.4~$M_{\sun}$
plotted as solid and dot-dashed lines, respectively.  The stellar birthline is plotted as a dotted line \citep{dan}. 
The filled symbols plot those sources with {\it Chandra} X-ray counterparts, 
the empty symbols, those without.    In Serpens, the six highly luminous objects 
without X-rays are very likely contaminating AGB stars, and are marked with asterisks.   
The cross indicates a typical uncertainty in the spectral types of 1.5 subclasses, from which we 
calculate representative formal uncertainties of  0.2 in $log({L_{\*}}/{L_{\sun}})$  and 0.02 in $log(T_{eff})$.  
}
\label{hrdscx}
\end{figure}

\begin{figure}
\epsscale{0.8}
\plotone{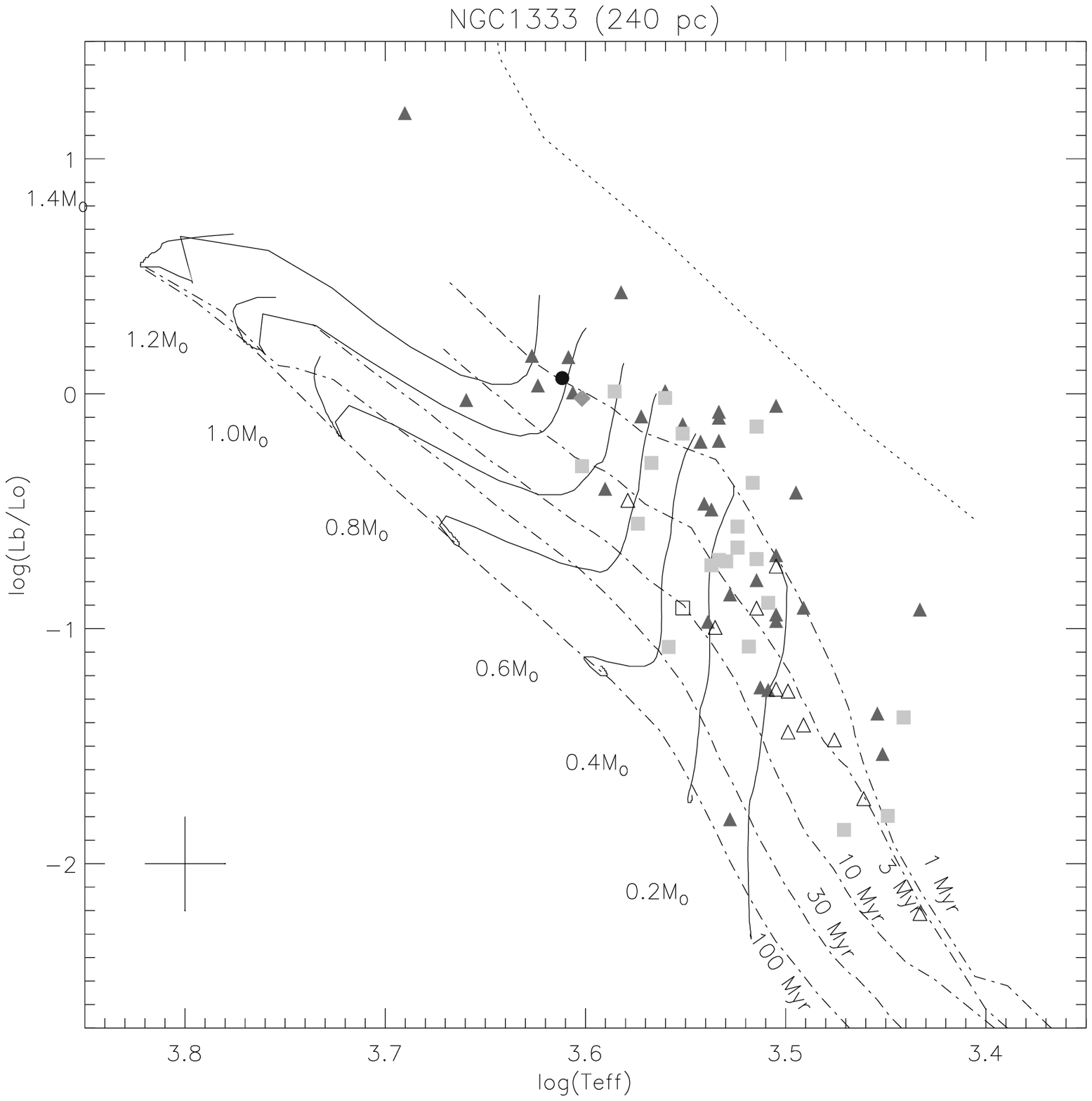}
\caption[HR diagram: NGC 1333 by class \& X-ray detection]{ 
The H-R diagram of the NGC 1333 cluster plotting all good spectral observations.  
The sources are identified by their evolutionary classification, with solid circles indicating 
the class I protostars, the inverted triangles the flat spectrum.   The Class II objects are 
represented by triangles, transition disks by open diamonds, and Class III sources 
by squares.    The evolutionary tracks are taken from \citet{bar}, with the 
1, 3, 10, 30, and 100~Myr isochrones and mass tracks from 0.2 to 1.4~$M_{\sun}$  
plotted as solid and dot-dashed lines, respectively.  The stellar birthline is plotted as a dotted line \citep{dan}.    
The filled symbols plot those sources with {\it Chandra} counterparts, the empty symbols  
those without.   There are no obvious candidates for AGB contamination in the NGC 1333 data. 
The cross indicates a typical uncertainty in the spectral types of 1.5 subclasses, from which we 
calculate representative formal uncertainties of  0.2 in $log({L_{\*}}/{L_{\sun}})$  and 0.02 in $log(T_{eff})$.  
}
\label{hrdncx}
\end{figure}

\begin{figure}
\epsscale{1.}
\plottwo{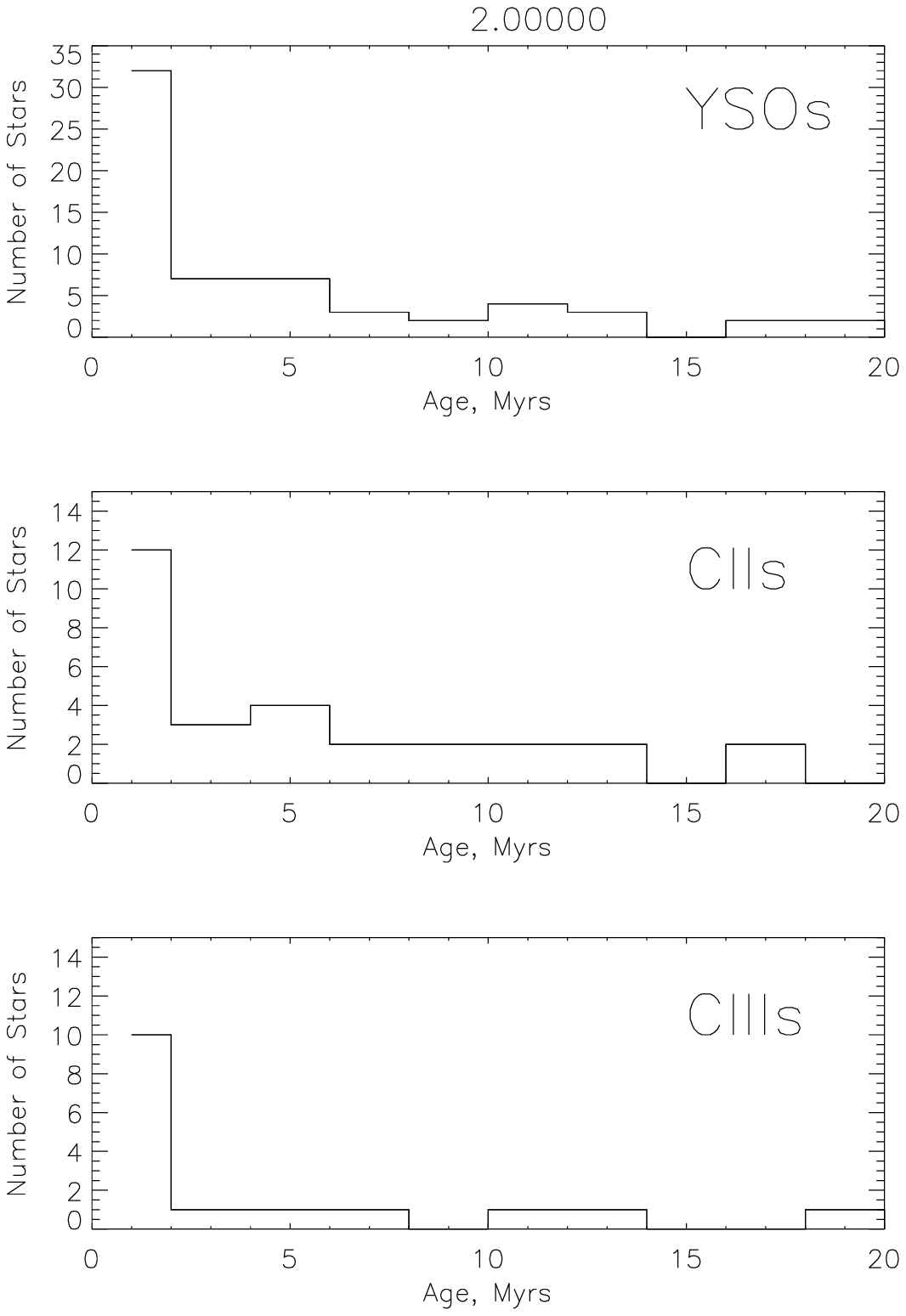}{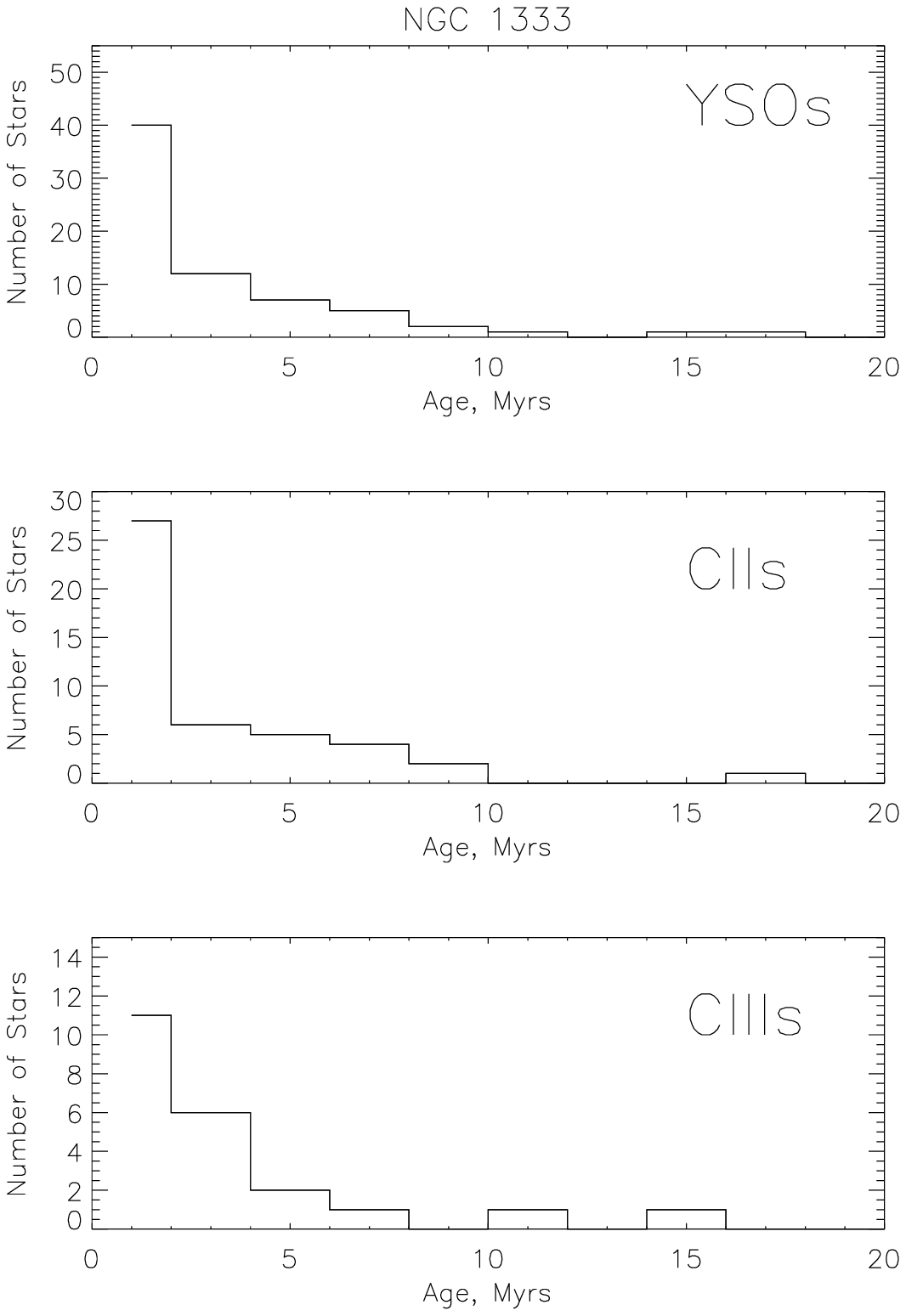}
\caption[Histogram of Ages in Serpens \& NGC 1333]{ 
The histograms of the ages of the Serpens ({\it left}) and NGC 1333 ({\it right}) young 
stars are presented.  The top plots show the histogram of all identified pre-main sequence 
stars, the central plot shows the Class II members only, and the lower plots show the 
Class III members only.  The distrubutions of number with age are similar between the 
three classes of objects in each cluster. 
}
\label{histoagesn}
\end{figure}

A spread in isochronal ages is observed in the HR diagrams of Serpens
and NGC 1333 (c.f. Figs.~\ref{hrdscx} \& \ref{hrdncx}).  Although the ages 
depend on the adopted pre-main sequence tracks, the presence of an age 
spread does not depend on the selected tracks.  The reason for this is that 
the spread in ages results from the large spread in luminosities apparent in the
HR-diagrams: for a given spectral type we see an order of magnitude
variation in the luminosities of the members.  For the remainder of
this discussion, we will use the ages derived from the \citet{bar}
tracks. Figure~\ref{histoagesn} presents the histograms of the ages of
the young stellar members for each cluster. In both clusters, a
similar distribution of ages is apparent: a peak at $< 2$~Myr and a
tail that extends to 10~Myr and beyond.  The shape of the inferred age
distribution is typical for age distributions derived from isochronal
studies of other young, embedded clusters
\citep{herbigic348,palla2004}.  Two explanations for the apparent
age distribution have emerged in the literature.  The first is that
this distribution indicates an accelerating star formation rate
\citep{pall}. The second is that the tail of the
distribution is largely the result of uncertainties in the
luminosities of stars or due to contamination from  stars external to the 
clusters or older populations of stars in the same region \citep{hart}.  
In Serpens, $41/66$ (62\%) of spectral detections are $<$3~Myrs old, with 
$25/66$ (38\%) older.  In NGC 1333, $51/74$ (68\%) of sources are $<$3~Mrs 
old, while $23/74$ (32\%) have ages above 3~Myrs.   To
further our understanding of the apparent age spreads, their validity,
and their origin, we examine how other properties of the stars depend
on their apparent isochronal ages.

The sources in the HR diagrams (Figs~\ref{hrdscx} \& \ref{hrdncx}) are 
distinguished by their evolutionary classification using the standard 
scheme for YSOs; a discussion of the methodology used to obtain these 
classifications is described in \citet{winston}.
The luminosities, and hence ages, of the protostellar Class I and Flat 
Spectrum sources can be affected by both
absorption and scattering light in infalling envelopes; consequently,
their ages are not considered to be reliable and we do not include
them in the following analysis.  The distribution of ages
of the Class III (stars without mid-IR excesses due
optically thick disks) and Class II (sources with mid-IR excesses due
to disks) show similar distributions for both clusters
(Figure~\ref{histoagesn}).  In Serpens the Class III sources have a
median age of $\sim$1~Myrs, while the Class II objects have an older
median age of $\sim$2.5~Myrs.  However, this difference is not
statistically significant: a KS test of the cumulative distribution
of the Class II and Class III ages resulting a probability of 38\%
that they were derived from the same parent distribution.  In NGC 1333
the median ages for the Class II and Class III members are 2.1~Myrs,
and 2.4~Myrs, respectively, with a similar K-S probability of 58\% of their
having the same parent distribution.  Thus, we find no evidence of different
age distributions between the Class II and Class III sources.

\begin{figure}
\epsscale{1.}
\plotone{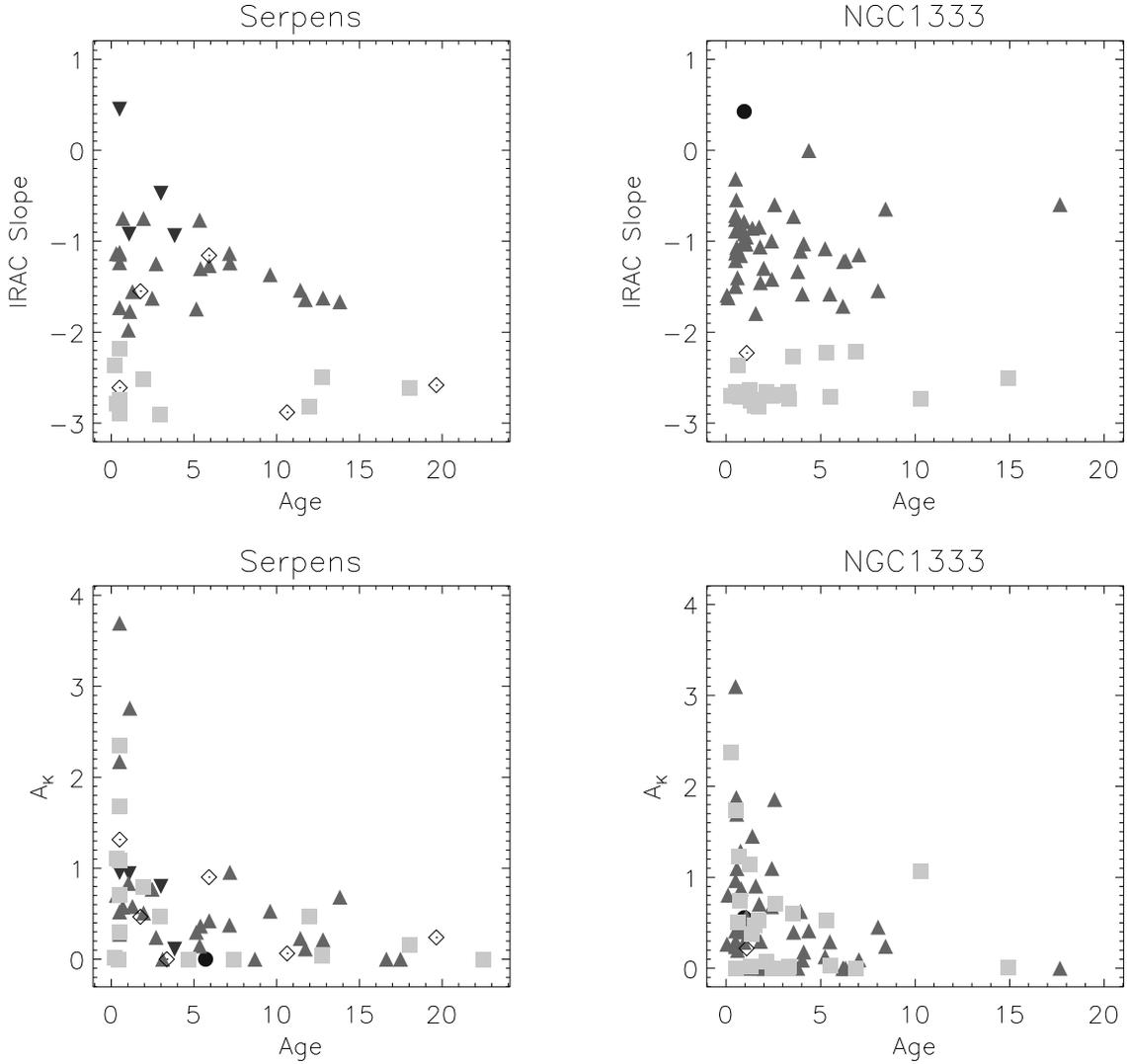}
\caption[Extinction with Age \& SED Slope for YSOs]{
$Above:$  The slope of the Spectral Energy Distribution (SED) calculated over the 
$IRAC$ bands, against stellar age.   There is an indication that the maximum slope 
of the Class II SEDs decrease with increasing age.   The photometry was dereddened 
before calculating the slope, to remove the effect of interstellar extinction.
$Below:$  The $K$-band extinction with calculated stellar age for Serpens (left) and 
NGC 1333 (right).  In both clusters, the younger objects possess higher values of 
extinction and are more thus deeply embedded in the cluster. 
The symbols indicate evolutionary class: Class 0/I: filled circles, flat spectrum: 
inverted triangles, Class II: triangles, Transition Disk: diamonds, Class III: squares.  
}
\label{akage}
\end{figure}

In Fig.~\ref{akage}, we examine both the spectral energy distribution
slope and extinction of the YSOs as a function of their apparent
isochronal ages.  The upper two panels in Fig.~\ref{akage} show
slopes of the spectral energy distributions (SEDs) determined with the
dereddened $IRAC$ (3-8~$\mu$m) photometry, $_{dered}\alpha_{IRAC}$.
The slope of the SED can be a function of the evolution of the disks
with lower slopes being associated with more evolved disks
\citep{lada,saguilar,jesusa}.  A range of slopes is observed
for the sources with ages $<$ 2 Myr.  For older sources, the upper
envelope of the observed slopes of Class II sources appears to
decrease with age in Serpens.  There is some indication of a similar
trend in NGC 1333.  Although these trends are tentative given the
small number of sources, they are similar to trends found in 
studies of clusters and associations with ages from 1-10~Myr.
These studies show the IRAC 3-8~$\mu$m slopes ($d log(\lambda F_{\lambda})
d log(\lambda)$) of Class II sources decrease with increasing age, with
the median slopes of $\sim 5-10$~Myr regions ranging from -1.5 to
-2 \citep{jesusb}.

These same studies show that the fraction of young stars with
IR-excesses from disks (i.e. the fraction of Class II and/or
transition disk sources) changes rapidly with the average age of a
cluster or association, decreasing from $80\%$ at 1 Myr, to $50\%$ at
3~Myr, and finally to $20\%$ at ages of 5-10~Myr \citep{jesusb}.
Interestingly, in NGC 1333 and Serpens, the fraction of sources with
disks does not follow this trend, although the statistics are poor.   
To examine the disk statistics, we use only the sources with X-ray
detections, these provide a relatively unbiased selection of pre-main
sequence stars with and without disks \citep{jesusa,preibisch,winston}; 
although, only a fraction of pre-main sequence stars with 
disks in Serpens and NGC 1333 are detected in X-rays \citep{win08b}.
The fraction of spectrally classified YSOs with X-ray detections in Serpens 
$<$3~Myrs is $20/37$ (54\%) and $>$3~Myrs is $10/29$ (34\%).  
In NGC 1333, the fractions are $39/52$ (75\%) for those $<$3~Myrs, and 
$14/24$ (58\%) for those older.  
For the pre-main sequence stars with X-ray detections
in the Serpens cluster, $9/20$ ($45\%$) of the $< 3$~Myr have disks
sources, while $4/10$ ($40\%$) of the sources $> 3$~Myr have disks.
In NGC 1333, $26/39$ (66\%) of the X-ray detected pre-main sequence
stars with ages $ <3$~Myr have disks while 7/14 ($50\%$) of the
sources with ages $> 3$~Myr have disks. Thus, the fraction of X-ray
luminous sources with disks does not decrease measurably with age. 
We also examined the fraction of sources with $H{\alpha}$ emission 
indicative of accretion and of $Li~I$ absorption in each of the two age groups.    
For Serpens and NGC 1333, respectively, the fractions for $H{\alpha}$ emission 
are $5/11$ (45\%) and $8/20$ (40\%) for the younger and $6/11$ (54\%) and 
$12/20$ (60\%) for the older sources. 
For $Li~I$ absorption the fractions are $2/12$ (17\%) and $18/32$ (56\%) for 
the younger sources and $10/12$ (83\%) and $14/32$ (44\%) for the older 
stars for the Serpens and NGC 1333 clusters, respectively.  
We find no statistical evidence of a difference in $H{\alpha}$ accretion between the 
younger and older populations in either cluster. There is no difference in the Lithium 
detection fraction in NGC 1333.  The lower detection fraction of $Li~I$ in younger 
Serpens sources is likely due to their higher extinction.

There have been suggestions in the literature that some of the apparently
older ($> 3$ Myr) objects are actually younger sources with nearly
edge-on disks \citep{peterson}.  In this case, the luminosity of the
source is attenuated by the disk; however, a contribution of scattered
light from the disk could result in an anomalously blue color and consequently,
an underestimate of the extinction and the stellar luminosity. 
A higher disk inclination may also lead to a lower 3-8~$\mu$m SED slope and
potentially explain the tentative dependence of observed age and slope
discussed above \citet{jesusa,jesusb}.  We find this explanation unlikely 
for two reasons.  
First, this effect cannot explain the older Class III objects: the Class II and 
Class III objects show similar age distributions.  Second, models of near 
edge-on disks show a steep dip in emission around 8~$\mu$m  \citep{dal};  
inspection of the SEDs shows no drop at 8~$\mu$m indicative of such a dip.  
Recent IRS spectroscopy of some of these sources will address the possibility
of edge on disks more definitively in the near future.

\begin{figure}
\epsscale{1.}
\plotone{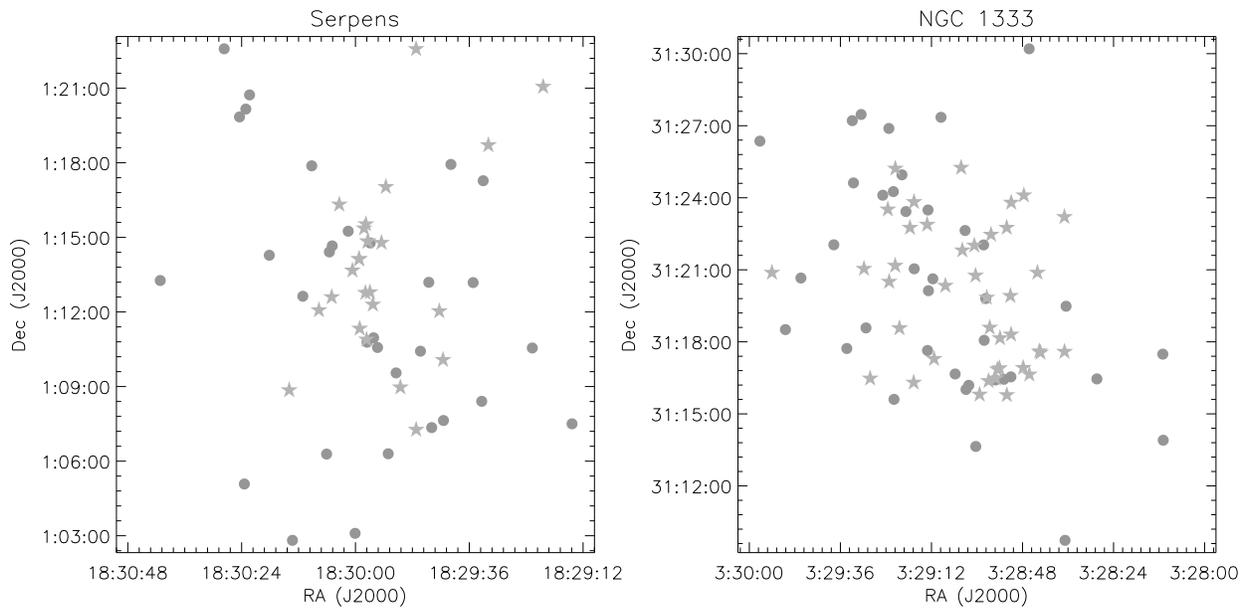}
\caption[Spatial distribution with Age for YSOs]{
The spatial distributions of the YSOs with age for Serpens (left) and NGC 1333 (right). 
The stars represent cluster members with ages less than 3~Myrs, the filled circles those 
with ages greater than 3~Myrs.   The younger objects in both clusters appear to be 
more centrally concentrated.   
}
\label{agespatdist}
\end{figure}

There is a strong dependence of apparent age on extinction and spatial
distribution.  In Figure~\ref{akage} the $K$-band extinctions, $A_K$,
in each of the clusters, is plotted against the apparent isochronal
age.  The youngest objects show the highest extinctions; objects with 
ages $< 3$ Myr have extinctions up to 4~$A_K$ while objects with 
ages $> 3$~Myr have extinctions less than 1~$A_K$.  There is also 
a corresponding difference in the spatial distribution as a function of 
age.  In order to examine the spatial distribution, the ages were 
binned into two categories: those objects with ages $<$3~Myrs  
and those $>$3~Myrs. Figure~\ref{agespatdist} shows the spatial 
distribution of cluster members with age for each cluster.  The 
nearest neighbour distances of the two age groups were found and compared with 
that of a random distribution over the spatial extent of the cluster, 
using the K-S test with $10^{3}$ iterations.  
In both clusters, the older stars appear to follow the random distributions,
with probabilities of 28\% for Serpens and 26\% for NGC 1333 of being 
drawn from the random distributions.  The younger members are more 
highly clustered, with probabilities of being drawn from the random distribution 
of much less than 1\% in each case.   It should be noted, that older members 
are found toward the central cluster, but do not show the same degree of concentration.  
Given the projected spatial extent of both clusters (about 1.5~pc) and random 
stellar velocities of $\sim$0.5~kms$^{-1}$, it would be possible for the older 
population to have originated in the more clustered regions and to have 
dispersed to their current locations.

Our results confirm the presence of stars with isochronal ages of
3-10~Myr toward these clusters.  We find it unlikely that the older
sources are contamination from galactic or extragalactic sources.
There was no trend in the detection with age in either $Li~I$ or $H\alpha$. 
The surface gravities of sources with ages $>$~3 and $<$~3 Myrs was examined, 
and again no trend was found in either cluster (Figure~\ref{surfgrav}). 
Furthermore, uncertainties in the extinction correction would primarily 
affect the sources with high extinctions, yet these show a lower dispersion in age. 
Finally, there is also no indication that the older sources have exceptionally
large uncertainties in their luminosity or temperature determinations. 
In general, if the age spread is due to the uncertainties or variations 
in the luminosity, it needs to be understood why these variation are more 
significant for the less reddened sources. 
Thus, toward both clusters we detect a young ($<$ 3 Myr) embedded population 
surrounded by a dispersed population of stars with apparent isochronal ages 
of 3-10 Myr. 

The data thus suggest that the regions of ongoing star formation, i.e. the embedded
clusters, are surrounded by a 3-10 Myr population of stars.   This could result from a 
previous episode of star formation in the vicinity of the cluster, or possibly, from stars 
that formed within the cluster that have since drifted out of the cloud.  
However, an interesting conundrum for this interpretation is the observed slope and 
the frequency of disks.   In support of the older ages for the dispersed young stars, 
there is an indication that the slope of the SEDs decreases with the isochronal age, 
consistent with disk evolution. However, we do not see a corresponding decrease 
in the frequency of disks that we would expect with age; this would argue that the
dispersed and embedded population have similar ages.   

Although this study has strengthened the case that the apparently older stars are 
real members of the observed star formation regions, further study is needed to address
potential biases in our study and further examine the nature of the $> 3$ Myr stars. 
One bias that could affect our arguments  is that older objects are fainter; 
consequently, deeply embedded, older objects would have not been selected for 
our spectroscopic observations.  This may have resulted in the lack of a detected age 
spread  for the more embedded stars.   This bias can be examined by spectroscopy of
 fainter embedded sources in these clouds.   Further spectroscopy of the apparently older 
 sources, both in the infrared and optical, may help in confirming the observed properties 
 (surface gravities, temperature, luminosity) as well as examining the properties of their disk.   
In conclusion, the apparent age spreads remain an important and difficult problem 
in understanding the star formation history of embedded clusters.

\section{\bf Conclusion}

This work presents the results of a far-red and near-IR spectroscopic survey of 
young stellar objects in two young clusters: the Serpens Cloud Core and NGC 1333.
The far-red spectra were obtained with the $Hectospec$ spectrograph on the MMT.  
Near-IR spectra were obtained with the $SpeX$ spectrograph and imager on the NASA IRTF.  
The spectral data were combined with previous {\it Spitzer IRAC} and {\it Chandra} 
studies of the regions to examine evolutionary class and X-ray detection as a function 
of physical age and stellar mass.

\begin{itemize}

\item  Spectral types were determined for 61 previously identified YSOs in the Serpens 
cluster, 30 with SpeX and 36 with Hectospec.  Of these, 36 had detections in the 
{\it Chandra} X-ray data.   In the NGC 1333 cluster, 69 spectral types were determined, 
22 with SpeX and 47 with Hectospec,  55 of which had detections in the X-ray data.  
The majority of YSOs with spectral types were Class II and Class III objects, the more 
deeply embedded protostars are more difficult to accurately type as the emission 
from their envelopes can alter the features of their spectra. 

\item Four possible new cluster members were identified by the presence of $Li~I$ 
in their spectra in Serpens, and five in NGC 1333. These would form part of the group 
of Class III objects not detected in the {\it Chandra} surveys of the regions.  

\item The consistency between spectral typing in each wavelength regime was 
tested by observing 6 stars in Serpens and 8 in NGC 1333 with both instruments.  
In all cases, the difference in spectral classification was less than 0.8 subclasses, with the 
majority less than 0.5 subclasses.  

\item  For the far-red spectra, the equivalent width of $H{\alpha}$ has been examined as a 
function of  spectral classification and evolutionary class.  Disk-bearing stars (Class II) exhibit 
higher $H{\alpha}$ emission than those without (Class III).  The Class III sources were 
found to consistently show no evidence of accretion, while the Class II sources generally do.   
Emission lines of $Pa{\beta}$, $Br{\gamma}$, $Ca~II$ and $O~I$ were also observed 
toward many of the YSOs.

\item   The HR-diagrams for both clusters were plotted to determine the ages and 
masses of the members and ascertain the age and star formation history for each cluster.  
The distribution of apparent isochronal ages show peaks at ages $<$~2 Myr and
tails extending out to 10 Myr and beyond. 

\item   In Serpens, the Class II objects were found to have an average age of  
$\sim$2.5~Myrs,  the Class III sources, an average age of $\sim$1~Myrs. 
In NGC 1333, the average isochronal ages of the Class II (2.1~Myrs) and 
Class III  (2.4~Myrs) sources were statistically similar. 
K-S tests give probabilities of 38\% and 58\% for Serpens and NGC 1333, respectively, 
of the ages of the Class II and III sources arising from the same distribution.     

\item  The spatial distribution and extinction of the spectrally identified YSOs was 
examined as a function of the apparent isochronal age.  YSOs with ages $<$3~Myr are 
found to be more deeply embedded and more clustered than the YSOs with ages $>$ 3~Myr. 
This suggests that the apparently older stars may be a more spatially extended  
component from previous star formation episodes.

\item  There is an indication that the slope of the SEDs of the Class II objects 
decreases with apparent isochronal age; consistent with other observations of disk 
evolution.  However, the disk fraction for the sample of X-ray detected objects
was found to remain constant with isochronal age.  It is not clear how we reconcile 
the isochronal ages with the high disk fractions. 

\item   It is unclear whether the apparent age spreads reflect actual age spreads as proposed 
in accelerating star formation scenarios,  or whether they result from overestimates for the ages 
of the pre-main sequence stars.   In support of actual age spreads,  the $>$3~Myr sources show 
less extinction, are not centrally condensed and have a hint of lower SED slopes for the Class II 
objects.   These suggest a previous episode of star formation or perhaps the drifting and dispersal  
of older stars from the embedded clusters.   However,  it is difficult to reconcile the older ages with 
the high percentage of disks found around the $>$3~Myr stars.   High dispersion spectroscopy of 
these objects should provide the means to compare the radial velocities and surface gravities of 
the $<$3~Myr and $>$3~Myr stars, and may help determine the origin of the apparent age spreads.

\end{itemize}

Visiting Astronomer at the Infrared Telescope Facility, which is operated by the University of Hawaii 
under Cooperative Agreement no. NCC 5-538 with the National Aeronautics and Space Administration, 
Science Mission Directorate, Planetary Astronomy Program.
This work is based on observations taken with the {\it Hectospec} instrument on the MMT, a joint 
venture of the Smithsonian Institute and the University of Arizona.  
This work is based on observations made with the {\it Spitzer} Space Telescope (PID 6, PID 174), 
which is operated by the Jet Propulsion Laboratory, California Institute of Technology under NASA 
contract 1407. Support for this work was provided by NASA through contract 1256790 issued by 
JPL/Caltech. Support for the IRAC instrument was provided by NASA through contract 960541 issued 
by JPL.
This publication makes use of data products from the Two Micron All Sky Survey, which is a 
joint project of the University of Massachusetts and the Infrared Processing and Analysis 
Center/California Institute of Technology, funded by the National Aeronautics and Space 
Administration and the National Science Foundation.
This research has made use of the NASA/IPAC Infrared Science Archive, which is operated by 
the Jet Propulsion Laboratory, California Institute of Technology, under contract with the 
National Aeronautics and Space Administration.
SJW was supported by NASA contract NAS8-03060.  This represents the results of VLA program S60872.
E. Winston would like to thank the Irish Research Council for Science, Engineering, 
and Technology (IRCSET) for funding and support.

\clearpage

\begin{center}
\begin{deluxetable}{ccc}
\tablecolumns{3}
\tabletypesize{\scriptsize}
\tablewidth{0pt}
\tablecaption{Fraction of previously identified YSOs, by cluster, for which we obtained spectral types \protect\label{ysonums}. }
\tablehead{  \colhead{Cluster}  & \colhead{Total} & \colhead{Spectra}   \\  }
\startdata
Serpens          &  138  &  62 (45\%)  \\   
Serpens X-ray    &   60  &  36 (58\%)  \\   
NGC 1333         &  169  &  69 (41\%)  \\   
NGC 1333 X-ray   &   86  &  55 (64\%)  \\   
\enddata
\end{deluxetable}
\end{center}

\begin{center}
\begin{deluxetable}{cccccccccc}
\tablecolumns{8}
\tabletypesize{\scriptsize}
\tablewidth{0pt}
\tablecaption{Serpens Hectospec sources \protect\label{hectotableS}}
\tablehead{
\colhead{ID\,\tablenotemark{a}}  & \colhead{SpT} & \colhead{Class\,\tablenotemark{b}}  &  \colhead{CID}  & \colhead{$T_{eff}$} & \colhead{$log(\frac{L_{bol}}{L_{\sun}})$} & \colhead{E.W. $H{\alpha}$} & \colhead{E.W. $Li~I$} & \colhead{$O~I$} & \colhead{$Ca~II$}   
}
\startdata

9  &      K8.0$\pm$1.5      &  I  &          67  &      3892  &    -0.462 &    -7.0 &    0.3  & d & \nodata \\
27  &      K7.0$\pm$3.5      &  FS   &        74  &      3997  &   -0.016   &      -58.1  &     \nodata  & \nodata & d  \\
33  &      M4.5$\pm$1.0      &  FS     &  \nodata  &       3197  &    -1.149    &  -17.9    &     0.2  & d & \nodata \\
36  &      K6.5$\pm$2.5      &  FS   &   9  &    4147  &     0.571    &     -43.5    &     \nodata  & d & d   \\
40  &      M0.0$\pm$1.5      &  FS   &   44  &   3821  &    -0.373     &    -140.4    &    \nodata   & \nodata & d     \\
53  &      M4.0$\pm$2.0     &  II   &  \nodata  &   3299  &   -1.564   &      -47.7    &   \nodata  & \nodata & \nodata   \\
54  &      M9.0$\pm$5.0    & II   &  \nodata  &    2400  &     -2.061   &      \nodata    &     \nodata  & \nodata & \nodata  \\
59  &      K7.5$\pm$1.5    &  II   &     77  &      3955  &    0.074    &     -48.78  &     \nodata   & d & \nodata    \\
66  &      M5.0$\pm$5.0    &  II   &  \nodata  &   3154  &     -2.235    &     -7.1  &     \nodata  & \nodata & \nodata  \\
67  &      M4.5$\pm$1.5    &  II   &  \nodata  &   3226  &    -1.502   &   -100.9     &   0.4  & d & d   \\
68  &      M8.7$\pm$5.0     &  II   &  \nodata  &    2493  &     -1.956  &        \nodata     &     \nodata  & \nodata & \nodata\\
71  &      M9.0$\pm$2.5      &  II   &  \nodata  &    2400    &   1.786    &    \nodata   &       \nodata  & d & \nodata  \\  
74  &      M4.0$\pm$1.5      &  II   &  \nodata  &   3241  &   -1.161   &      -8.1   &    \nodata   & \nodata & \nodata   \\
82  &      A7.0$\pm$2.5      &  II   &  \nodata  &   7850  &    0.846  &    11.1 &    \nodata  & \nodata & \nodata     \\
88  &      M7.5$\pm$1.5      &  II   &  \nodata  &   2778  &   -1.834   &    -49.9   &   \nodata   & d & \nodata \\
91  &      M5.5$\pm$1.5      &  II   &  \nodata  &    3098  &   -1.974    &   \nodata   &   \nodata  & \nodata & \nodata   \\
94  &      M1.5$\pm$1.0    &  II   &    57  &      3647  &    -0.713   &    -30.8   &   \nodata   & \nodata & \nodata   \\
96  &      M4.5$\pm$2.5      &  II    &  \nodata  &    3226.5   &  -0.983   &  -44.2  &    \nodata   & \nodata & \nodata \\
101  &      M4.0$\pm$1.0      &  II     &    87  &    3255  &    -1.143   &    -14.1  &   0.3   & d & \nodata  \\
102  &      M5.0$\pm$5.0    &  II   &  \nodata  &    3111  &   -2.385   &    \nodata    &   \nodata   & \nodata & \nodata \\
103  &      K2.5$\pm$1.5     &  II    &  \nodata  &   4832  &   0.138  &   -43.1    &    0.5   & d & d   \\
104  &      M0.5$\pm$1.5    &  II     &  \nodata  &    3763  &    -0.145   &      -11.8   &    0.2    & \nodata & \nodata    \\
124  &      K6.5$\pm$4.5 &      TD    &  \nodata  &    4161  &   -0.192    &    12.2    &  \nodata  & \nodata & \nodata  \\
128  &      M3.5$\pm$1.5   &    TD   &      46  &      3371  &    -1.217    &     -3.5  &   0.4   & \nodata & \nodata  \\
156  &      M2.0$\pm$5.0  &   TD    &  \nodata  &   3589  &    -0.704    &     -62.4    &   \nodata   & d & \nodata   \\
157  &      M1.0$\pm$3.5  &      TD    &    45  &    3690  &     -2.631    &     -8.7   &     0.4  & d & d   \\
166  &      M5.0$\pm$1.0   &   TD   &  \nodata  &    3125  &     -1.062   &      -14.2    &    0.3    & \nodata & \nodata     \\
190  &      M9.0$\pm$2.0      &  III   &        88  &      2400  &    -0.349   &    -5.9 &    \nodata   & \nodata & \nodata  \\
201  &      M5.0$\pm$1.0   &  III   &     62  &      3125  &     -1.592    &     -9.8    &     0.7   & \nodata & \nodata  \\
203  &      M5.0$\pm$1.5    &  III   &    51  &      3125  &     -1.997    &     -11.3   &     0.4    & \nodata & \nodata   \\
204  &      M8.5$\pm$5.0    &  III   &     65  &      2524  &     -1.864   &     \nodata    &     \nodata   & \nodata & \nodata \\
225  &      G2.5$\pm$1.0   &  III   &    56  &      5851  &     0.406   &     -0.3    &      0.4   & \nodata & \nodata  \\
230  &      K5.0$\pm$2.5   &  \nodata    &  \nodata  &       4321  &    -0.319   &     -0.3  &    0.3    & \nodata & \nodata \\
232  &      M1.0$\pm$5.0  &  \nodata    &  \nodata  &       3705  &    -0.523    &     -2.9  &       0.5   & \nodata & \nodata   \\
233  &      M4.5$\pm$2.0  &  \nodata      &  \nodata  &    3197  &    -1.567    &   -8.4    &    0.6   & d & d \\
235  &      K6.5$\pm$4.0   &  \nodata      &  \nodata  &       4161  &   -0.701   &   2.1   &     0.4    & d & \nodata  \\

\enddata
\tablenotetext{a}{See Table\ref{tblsernewcIII} and Table~4 of \citet{winston} for coordinates and magnitudes associated with spectra. }
\tablenotetext{b}{I: Class I, FS: flat spectrum, II: Class II, TD: Transition Disk, III: X-ray Detected Class III. Those without a class are the new candidate Class IIIs with detected $Li~I$ absorption. }

\end{deluxetable}
\end{center}

\begin{center}
\begin{deluxetable}{cccccccccc}
\tablecolumns{8}
\tabletypesize{\scriptsize}
\tablewidth{0pt}
\tablecaption{NGC 1333 Hectospec sources \protect\label{hectotableN}}
\tablehead{
\colhead{ID\,\tablenotemark{a}}  & \colhead{SpT} & \colhead{Class\,\tablenotemark{b}}  &  \colhead{CID}  & \colhead{$T_{eff}$} & \colhead{$log(\frac{L_{bol}}{L_{\sun}})$}  & \colhead{E.W. $H{\alpha}$}  & \colhead{E.W. $Li~I$ }  & \colhead{$O~I$} & \colhead{$Ca~II$}    
}
\startdata
18    &   K7.0$\pm$1.0    &     I    &   98    &   4089    &   0.066   &    -35.8     &    \nodata    & d & d   \\
45    &   M5.0$\pm$1.0    &    II    &   97    &   3125    &   -0.420    &     -30.9   &     \nodata    & \nodata & \nodata   \\
46    &   M7.5$\pm$1.0    &    II    &    139    &   2829    &   -1.533   &    -16.1     &   \nodata    & \nodata & \nodata    \\
47    &  M3.0$\pm$1.5    &    II    &   93    &   3415    &   -0.102    &  -16.5   &   0.3       & d & \nodata      \\
49    &   M8.0$\pm$1.0    &    II    &   90    &   2710    &   -0.918    &    -199.6   &    \nodata     & d & d    \\
50    &   M4.5$\pm$1.0    &    II    &    84    &  3197.5    &   -0.052   &    -76.2    &    0.2    & d & d    \\
51    &   M4.0$\pm$1.0    &    II    &   133    &   3255.5    &   -1.250   &   -45.7    &      0.3   & \nodata & d  \\
52    &   K7.0$\pm$1.0    &   II    &    83    &   4039    &   0.005   &   -7.9   &    0.5     & \nodata & \nodata  \\
53    &   M7.0$\pm$2.0   &   II    &   81    &   2846    &   -1.361    &   -35.2   &   0.3       & \nodata & \nodata      \\
54    &   M2.5$\pm$1.5    &    II    &   139    &  3473    &   -0.467   &   -5.8   &   0.6      & \nodata & \nodata  \\
55    &   M5.0$\pm$1.0    &    II    &  \nodata &   3154    &    -1.440    &   -26.0    &   \nodata   & \nodata & \nodata   \\
58    &   M5.0$\pm$1.0    &   II    &    131    &   3098    &  -0.911     &   -34.5   &   0.4     & d & \nodata   \\
59    &   M1.0$\pm$2.5   &   II    &   78    &   3734    &   -0.096   &    -32.9   &   \nodata      & \nodata & \nodata     \\
62    &   M0.5$\pm$1.5    &    II    &  \nodata &   3792    &   -0.453   &    -176.7    &     0.4     & \nodata & d   \\
69    &   M3.5$\pm$1.5    &   II    &    70    &   3371.5    &   -0.854    &     -26.5   &   \nodata    & \nodata & \nodata   \\
76    &   M3.5$\pm$1.0    &    II    &    57    &  3371.5    &    -1.811    &    -5.5   &   \nodata     & \nodata & \nodata   \\
78    &   M2.5$\pm$1.0    &   II    &    54    &   3458.5  &  -0.970   &    -167.6    &   0.4      & d & d   \\
82    &   K6.0 $\pm$1.5    &   II    &    51    &   4234    &  0.161     &  -33.9    &   0.3       & \nodata & d   \\
94    &   M5.0$\pm$1.0    &  II    &  \nodata  &   3098    &  -1.410    &    -38.4      &    0.4     & d & \nodata  \\
104    &   M4.5$\pm$1.5    &   II    &  \nodata &   3154    &   -1.265    &   -58.1   &   \nodata     & \nodata & \nodata  \\
105    &   M8.0$\pm$1.0     &   II    &  \nodata &    2710    &  -2.212   &    -59.6   &    \nodata    & \nodata & \nodata  \\
109    &  M3.0$\pm$1.0    &   II    &   24    &   3444    &  -0.492    &    -28.8   &   0.1     & \nodata & \nodata   \\
111    &   M0.0 $\pm$1.5    &   II    &    22    &   3821    &    0.432    &    -39.3   &    0.5   & \nodata & \nodata  \\
114    &   M0.0$\pm$1.5    &   II    &   17    &  3892    &  -0.404    &   -9.8   &   0.6      & d & \nodata   \\
118    &   M4.5$\pm$1.5    &   II    &   9    &  3197.5    &  -0.967    &   -8.8   &  0.6      & d & \nodata \\
119    &   K4.0$\pm$3.0  &     II    &   161    &  4566    &  -0.026   &   -116.9    &   0.2     & d & d   \\
120    &   M4.5$\pm$1.5    &   II    &   113    &  3197.5    &   -0.939   &    -51.1   &      0.2   & d & \nodata   \\
121    &   M4.5$\pm$1.5    &   II    &  \nodata &  3197.5  &   -1.257   &   -18.8   &   0.2       & \nodata & \nodata  \\
125    &   M4.5$\pm$1.5    &   II    &   106    &  3197.5    &   -0.687    &  -3.5    &     0.4     & \nodata & \nodata  \\
127    &   M3.0$\pm$1.0    &   II    &  \nodata &   3429.5    &   -0.994    &   -142.2    &    0.2    & d & d  \\
128    &   M4.5$\pm$1.5    &   II    &    169    &  3226.5    &    -1.261   &   -5.2  &   0.5     & \nodata & \nodata \\
131    &   M7.0$\pm$1.5    &   II    &   \nodata &    2891    &  -1.723    &   -25.9  &   \nodata      & \nodata & \nodata   \\
133    &   M4.5$\pm$1.5    &   II    &   \nodata &   3197.5    &   -0.733   &   -106.1    &  0.1     & d & d   \\
136    &   K7.5$\pm$1.5    &  TD   &    157    &    3997    &  -0.019    &  -3.7    &   0.6     & \nodata & \nodata  \\
137    &   K7.5$\pm$1.5    &   III    &    2    &   3997    &   -0.308    &   -6.4    &    0.3    & \nodata & \nodata   \\
138    &   M4.0$\pm$1.0    &    III    &   104    &    3299    &  -1.076    &  -5.1   &   0.7      & \nodata & \nodata  \\
140    &   M1.0$\pm$1.0    &   III    &   102    &   3690.5    &   -0.294    &   -3.4   &    0.3     & \nodata & \nodata \\
150    &   M4.5$\pm$1.5    &   III    &    60    &    3226.5    &    -0.890   &    -8.3   &     \nodata    & \nodata & \nodata  \\
157    &  M7.5$\pm$1.5    &   III    &    33    &   2812    &   -1.796   &   -6.2    &  \nodata      & \nodata & \nodata  \\
160    &   M3.5$\pm$1.5    &   III    &   23    &   3342.5    &   -0.565    &   -4.0   &    0.4  & \nodata & \nodata \\
162    &   M3.5$\pm$1.0    &    III    &    8    &    3386    &  -0.713   &    -6.2    &   0.4     & \nodata & \nodata     \\
164    &   M7.5 $\pm$1.0    &   III    &   156    &  2761    &   -1.377   &     -26.7   &   0.1     & d & \nodata   \\
165    &   M3.0$\pm$1.5    &   III    &    154    &    3444    &  -0.730    &     -4.2    &   0.4    & \nodata & \nodata  \\
167    &   M3.0$\pm$1.5    &  \nodata &   \nodata &   3444    &  -0.626   &    -3.9   &   0.4     & \nodata & \nodata  \\
170    &   M3.5$\pm$1.0    &  \nodata &   \nodata &   3313.5    &   -2.116    &   -3.8    &   0.2    & \nodata & \nodata  \\
174    &   M4.5$\pm$1.5    &  \nodata &  \nodata &   3226.5    &  -0.968   &  -9.8    &   0.4      & \nodata & \nodata  \\
175    &   M3.0$\pm$1.0    &  \nodata &   \nodata &   3386    &   -1.782    &    -5.1    &   0.2     & \nodata & \nodata   \\  
176    &   M3.0$\pm$1.5    &  \nodata &   \nodata &   3429.5    &   -0.696    &    -2.4   &    0.5   & \nodata & \nodata  \\
177    &   M6.5 $\pm$1.5    &  III    &   107    &   2957    &   -1.856    &   -19.3    &   \nodata    & \nodata & \nodata   \\
178    &   M0.5$\pm$1.0    &   III   &    68    &   3748.5    &    -0.552   &   -2.6   &   0.4     & \nodata & \nodata   \\   
179    &   M1.5$\pm$1.0    &   III    &    55    &   3618    &   -1.077    &   -2.3   &    0.2      & \nodata & \nodata \\
180    &   M4.0$\pm$1.5    &   III    &    71    &   3284.5   &   -0.379    &   -8.2   &    \nodata    & d & \nodata  \\

\enddata
\tablenotetext{a}{See Table~7 in \citet{win08b} \& \citet{gutn} for the coordinates and magnitudes associated wih spectra. }
\tablenotetext{b}{I: Class I, FS: flat spectrum, II: Class II, TD: Transition Disk, III: X-ray Detected Class III. Those without a class are the new candidate Class IIIs with detected $Li~I$ absorption. }

\end{deluxetable}
\end{center}

\begin{center}
\begin{deluxetable}{cccccccc}
\tablecolumns{6}
\tabletypesize{\scriptsize}
\tablewidth{0pt}
\tablecaption{Serpens SpeX sources \protect\label{spextableS}}
\tablehead{
\colhead{ID\,\tablenotemark{a}}  & \colhead{SpT} & \colhead{Class\,\tablenotemark{b}}  &  \colhead{CID}  & \colhead{$T_{eff}$} & \colhead{$log(\frac{L_{bol}}{L_{\sun}})$}  & \colhead{Br$_\gamma$} & \colhead{Pa$_\beta$} 
}
\startdata

       59    &     M0$\pm$1.5        &      II    &    77    &     3850   &    0.049    & d & d    \\
       65  &       K7$\pm$1.5     &      II   &   \nodata    &     4060   &      -0.667  & d & d      \\
       73    &     M2.5$\pm$1.5     &     II    &     86  &       3487.5   &    -0.108   & \nodata & \nodata     \\
       78    &     M0$\pm$1.5         &    II   &      84    &     3850    &   -0.260    & \nodata & \nodata    \\
       79  &       M2$\pm$1.5     &      II   &   \nodata    &     3560   &      -0.967   & \nodata & \nodata      \\
       81     &    M2$\pm$1.5       &      II    &     60    &     3560  &     -0.322     & d & d    \\
       83     &      K2$\pm$1.5       &     II    &     71    &     4900   &     0.644    & d & d     \\ 
       84     &    M1$\pm$1.5         &     II    &     42    &     3705  &     -0.666     & \nodata & \nodata    \\
       85     &    M3$\pm$1.5        &     II     &    68    &     3270  &    0.124       & \nodata & \nodata  \\
       87     &    M8$\pm$1.5        &     III     &    61   &      2710   &     -1.666    & \nodata & \nodata     \\
       98  &       M5$\pm$1.5     &      II   &   85    &     3125   &      -1.191        & \nodata & \nodata \\  
     100   &     M4.5$\pm$1.5   &       II   &      59    &     3415.5   &    -0.891      & \nodata & \nodata   \\
     105   &      K1$\pm$1.5      &     II   &      73    &     5080    &     1.364       & \nodata & \nodata  \\
     122  &       M3.5$\pm$1.5     &      TD   &   \nodata    &     3342.5   &      1.412  & \nodata & \nodata       \\
     127   &     M2$\pm$1.5      &     TD    &   \nodata      &    3560   &     0.808       & \nodata & \nodata  \\
     134  &       M2$\pm$1.5     &      TD   &   \nodata     &     3560   &     0.921       & \nodata & \nodata  \\
     158  &       M1$\pm$1.5     &      TD   &   \nodata    &     3705   &      2.667       & \nodata & \nodata  \\
     170  &       M3.5$\pm$1.5     &    TD   &   \nodata    &     3342.5   &      0.849     & \nodata & \nodata    \\
     176  &       M4$\pm$1.5     &      TD   &   \nodata    &     3270   &      0.151       & \nodata & \nodata  \\
       190   &      M2$\pm$1.5      &     III    &     88    &     3560    &   -0.181      & \nodata & \nodata   \\
     192  &       A0     &      III   &   69    &     9520   &      1.602       & \nodata & \nodata  \\
       193  &       A0     &      III   &   50    &     9520   &      1.198      & \nodata & \nodata   \\
       196   &      M5$\pm$1.5      &      III    &     76    &     3125   &    -0.592     & \nodata & \nodata    \\
     199   &      M2.5$\pm$1.5     &    III   &      40    &     3487.5   &    0.026       & \nodata & \nodata  \\
       209   &      M4.5$\pm$1.5    &     III    &     41   &      3197.5   &    -0.955      & \nodata & \nodata   \\
     205  &       K5$\pm$1.5     &     III   &   64    &     4350   &      -0.657        & \nodata & \nodata \\
        215   &      M7$\pm$1.5       &   III    &     58    &     2880   &    -0.628     & \nodata & \nodata    \\
       216   &      M0.5$\pm$1.5    &    III   &     80    &     3777.5    &    0.415     & \nodata & \nodata    \\
      220   &     A3       &    III    &     36   &      8720  &    -0.060       & \nodata & \nodata  \\
        221  &       K7.5$\pm$1.5     &   III   &   37    &     3955   &      -0.290      & \nodata & \nodata   \\

\enddata
\tablenotetext{a}{See \citet{winston} for the associated coordinates and magnitudes of these sources. }
\tablenotetext{b}{I: Class I, FS: flat spectrum, II: Class II, TD: Transition Disk, III: Class III. }

\end{deluxetable}
\end{center}

\begin{center}
\begin{deluxetable}{cccccccc}
\tablecolumns{6}
\tabletypesize{\scriptsize}
\tablewidth{0pt}
\tablecaption{NGC 1333 SpeX sources \protect\label{spextableN}}
\tablehead{
\colhead{ID\,\tablenotemark{a}}  & \colhead{SpT} & \colhead{Class\,\tablenotemark{b}}  &  \colhead{CID}  & \colhead{$T_{eff}$} & \colhead{$log(\frac{L_{bol}}{L_{\sun}}$)} & \colhead{Br$_\gamma$\,\tablenotemark{c}} & \colhead{Pa$_\beta$\,\tablenotemark{c}} 
}
\startdata

       48    &       M4$\pm$1.5    &       II  &         91    &       3270    &     -0.793     & \nodata & d   \\
       57    &       K7$\pm$1.5    &      II  &         79    &       4060    &      0.156      & \nodata & \nodata  \\
       61    &       M3$\pm$1.5    &       II  &         76    &       3415    &     -0.200     & d & d   \\
       64    &       M2$\pm$1.5    &       II  &         74    &       3560    &     -0.129      & d & d  \\
       65    &       M1.5$\pm$1.5    &     II  &         73    &       3632.5   &     0.0111    & d & d    \\
       67    &       M2.5$\pm$1.5    &     II  &         72    &       3487.5  &     -0.204     & d & d   \\
       71    &       M6$\pm$1.5    &       II  &       \nodata    &         2990  &    -1.473   & \nodata & \nodata     \\
       73    &       K2$\pm$1.5    &       II  &         64    &       4900    &       1.195    & \nodata & \nodata    \\
       88    &       A3    &       II  &         46    &       8720    &       1.394      & \nodata & \nodata  \\
       91    &       M2.5$\pm$1.5    &     II  &         \nodata  &         3487.5  &    -0.802  & d & \nodata      \\
       101    &       M4$\pm$1.5    &      II  &       \nodata    &         3270  &    -0.913    & \nodata & \nodata    \\
       106    &       M3$\pm$1.5    &      II  &         31    &       3415    &    -0.077      & \nodata & \nodata  \\       
       116    &       K6$\pm$1.5    &      II  &         11    &       4205    &     0.035      & d & d  \\
       139    &       M0$\pm$1.5    &      III  &         103    &       3850    &    0.009     & \nodata & \nodata   \\
       143    &       M4$\pm$1.5    &      III  &         85    &       3270    &     -0.139    & \nodata & \nodata    \\
       149    &       M3$\pm$1.5    &      III  &         162    &       3415    &     -0.708   & \nodata & \nodata     \\
       151    &       M1.5$\pm$1.5    &    III  &         58    &       3632.5   &    -0.018     & \nodata & \nodata   \\
       154    &       M2$\pm$1.5    &      III  &         121    &       3560    &     -0.169    & \nodata & \nodata    \\
       156    &       M4$\pm$1.5    &      III  &         41    &       3270    &     -0.704     & \nodata & \nodata   \\
       159    &       M2$\pm$1.5    &      III  &       \nodata    &         3560  &    -0.912   & \nodata & \nodata     \\
       161     &       A0    &      III  &       \nodata    &         9520  &    1.405     & \nodata & \nodata   \\
       163    &       M3.5$\pm$1.5    &     III  &         6  &         3342.5  &    -0.655      & \nodata & d  \\

\enddata
\tablenotetext{a}{See \citet{gutn,win08b} for coordinates and magnitudes associated wih spectra. }
\tablenotetext{b}{I: Class I, FS: flat spectrum, II: Class II, TD: Transition Disk, III: Class III. }
\tablenotetext{c}{'d' indicates detection of this feature. }

\end{deluxetable}
\end{center}

\begin{center}
\begin{deluxetable}{ccccccccccccc}
\tablecolumns{13}
\tabletypesize{\scriptsize}
\setlength{\tabcolsep}{0.03in}
\rotate
\tablewidth{0pt}
\tablecaption{ Photometry for Serpens Candidate Class III YSOs \protect\label{tblsernewcIII}}
\tablehead{
\colhead{ID\,\tablenotemark{a}}  & \colhead{RA (2000)}   & \colhead{Dec (2000)} & \colhead{J}  & \colhead{H} 
 & \colhead{K} & \colhead{3.6${\mu}m$} & \colhead{4.5${\mu}m$}  & \colhead{5.8${\mu}m$}  & \colhead{8.0${\mu}m$}  & \colhead{24.0${\mu}m$}  &  \colhead{$\alpha_{IRAC}$\,\tablenotemark{b}}  & \colhead{A$_{K}$} 
}
\startdata

230  &  18:29:07.08  &  1:04:6.31  &  12.951$\pm$0.022  &  11.688$\pm$0.029  &  11.281$\pm$0.024  &  10.977$\pm$0.002  &  10.981$\pm$0.004  &  10.929$\pm$0.008  &  10.885$\pm$0.011  &  \nodata  &  -2.77  &  0.69    \\

232  &  18:29:22.56  &  1:07:03.88  &  12.980$\pm$0.024  &  11.815$\pm$0.031  &  11.438$\pm$0.021  &  11.126$\pm$0.003  &  11.081$\pm$0.003  &  11.012$\pm$0.007  &  11.013$\pm$0.008  &  \nodata  &  -2.75  &  0.59   \\

233  &  18:29:33.07  &  1:07:16.01  &  14.566$\pm$0.033  &  13.668$\pm$0.042  &  13.268$\pm$0.039  &  12.921$\pm$0.006  &  12.901$\pm$0.008  &  12.801$\pm$0.024  &  12.780$\pm$0.043  &  \nodata  &  -2.71  &  0.23   \\

235  &  18:29:39.77  &  1:20:41.14  &  13.677$\pm$0.027  &  12.471$\pm$0.032  &  11.962$\pm$0.021  &  11.606$\pm$0.003  &  11.647$\pm$0.004  &  11.651$\pm$0.012  &  11.509$\pm$0.016  &  \nodata  &  -2.78  &  0.61   \\

\enddata

\tablenotetext{a}{{\it Spitzer} Identifier. }
\tablenotetext{b}{Slope of SED calculated from available IRAC data points. }

\end{deluxetable}
\end{center}

\begin{center}
\begin{deluxetable}{ccccccccccccc}
\tablecolumns{13}
\tabletypesize{\scriptsize}
\setlength{\tabcolsep}{0.03in}
\rotate
\tablewidth{0pt}
\tablecaption{ Photometry for NGC 1333 Candidate Class III YSOs \protect\label{tbln13newcIII}}
\tablehead{
\colhead{ID\,\tablenotemark{a}}  & \colhead{RA (2000)}   & \colhead{Dec (2000)} & \colhead{J}  & \colhead{H}  & \colhead{K} & \colhead{3.6${\mu}m$} 
& \colhead{4.5${\mu}m$}  & \colhead{5.8${\mu}m$}  & \colhead{8.0${\mu}m$}  & \colhead{24.0${\mu}m$}   &  \colhead{$\alpha_{IRAC}$\,\tablenotemark{b}} & \colhead{A$_{K}$}
}

\startdata

167  &  3:28:11.01  &  31:17:29.26  &  12.435$\pm$0.019  &  11.399$\pm$0.028  &  10.998$\pm$0.025  &  10.746$\pm$0.002  &  10.662$\pm$0.003  &  10.605$\pm$0.006  &  10.589$\pm$0.008  &  \nodata  &  -2.71  &  0.34  \\

170  &  3:28:46.24  &  31:30:12.13  &  15.022$\pm$0.022  &  14.452$\pm$0.037  &  14.149$\pm$0.032  &  13.875$\pm$0.008  &  13.793$\pm$0.011  &  13.721$\pm$0.060  &  13.650$\pm$0.081  &  \nodata  &  -2.63  &  0.18   \\

174  &  3:29:46.39  &  31:20:39.52  &  12.351$\pm$0.019  &  11.732$\pm$0.022  &  11.474$\pm$0.022  &  11.199$\pm$0.003  &  11.122$\pm$0.004  &  11.033$\pm$0.008  &  11.051$\pm$0.013  &  \nodata  &  -2.71  &  0.10   \\

175  &  3:29:50.48  &  31:18:30.61  &  14.355$\pm$0.022  &  13.721$\pm$0.026  &  13.492$\pm$0.027  &  13.313$\pm$0.008  &  13.226$\pm$0.010  &  13.115$\pm$0.048  &  13.061$\pm$0.071  &  \nodata  &  -2.59  &  0.05   \\

176  &  3:29:57.22  &  31:26:21.47  &  12.354$\pm$0.02  &  11.537$\pm$0.026  &  11.243$\pm$0.019  &  11.059$\pm$0.002  &  10.987$\pm$0.003  &  10.904$\pm$0.008  &  10.913$\pm$0.009  &  \nodata  &  -2.72  &  0.26  \\

\enddata

\tablenotetext{a}{{\it Spitzer} Identifier.  }
\tablenotetext{b}{Slope of SED calculated from available IRAC data points.  }

\end{deluxetable}
\end{center}

\begin{center}
\begin{deluxetable}{ccccccc}
\tablecolumns{7}
\tabletypesize{\scriptsize}
\tablewidth{0pt}
\tablecaption{Evolutionary Classification of the Spectral Sources \protect\label{goodspecclass}. }
\tablehead{
\colhead{Cluster}  & \colhead{I} & \colhead{FS}  &  \colhead{II}  & \colhead{TD} & \colhead{III}  & \colhead{[III]}  \\
}
\startdata
Serpens             &  1 & 4  & 27  & 12  & 17 & 4 \\   
Serpens X-ray  & 1  & 3  & 12  & 2  & 17 & 0 \\   
NGC1333  & 1  & 0  & 45  & 1  & 22 & 5 \\   
NGC1333 X-ray &  1 & 0  & 33  & 1  & 20 & 0 \\   
\enddata
\end{deluxetable}
\end{center}




\clearpage

\end{document}